\documentclass{statsoc}
\usepackage[a4paper]{geometry}

\usepackage{natbib,float}
\usepackage{pgf}
\usepackage{tikz}
\usetikzlibrary {arrows,decorations}
\usepackage{pgfplots}
\pgfplotsset{compat=1.15}

\usepackage{amsmath}
\usepackage{amssymb}
\usepackage{ifthen}

\usepackage{times}
\usepackage[T1]{fontenc}

\usepackage{url}

\newcommand{\E}{\mbox{E}}

\title[Multilevel Longitudinal Social Networks]{Multilevel Longitudinal Analysis
of Social Networks}

\author[Johan Koskinen and Tom A.B. Snijders]{Johan Koskinen}
\address{University of Melbourne,
         Melbourne,
                  Australia.}
\email{jkoskinen@unimelb.edu.au}                  
\author[Johan Koskinen and Tom A.B. Snijders]{Tom A.B. Snijders}
\address{University of Oxford,
            Oxford,
            United Kingdom; 
            University of Groningen,             
            Groningen,             
            The Netherlands.}
\email{tom.snijders@nuffield.ox.ac.uk}

\newcommand{\motif}[7]{\footnotesize\begin{tikzpicture}
\tikzstyle{every node}=[x=3ex,y=3ex,inner sep=0pt,minimum size=4ex,shape=circle,draw,
                fill=none,font=\bfseries]
\tikzstyle{every path}=[shorten <=2pt,shorten >=2pt, ->, ultra thick,-stealth,
            decoration={snake,amplitude=0.5ex,segment length=2ex,post length=1ex}]
\ifthenelse{\isodd{#7}}{\node (i) at (0,0)  {\textbf{\textit{i}}}}
                {\node (i) at (0,0)  {\textbf{\textit{j}}}};
\ifthenelse{\isodd{#7}}{\node (j) at (4,0)   {\textbf{\textit{j}}}}
            {\node (j) at (4,0)   {\textbf{\textit{i}}}};
\node[shape=rectangle] (h) at (#1,#2)  {\textbf{\textit{h}}};
\ifthenelse{\isodd{#6}}{\node[shape=rectangle] (l) at (8,0)  {$\mathbf{\ell}$}}{};
\ifthenelse{\isodd{#3}}{\draw (i) -> (j)}{\draw (j) -> (i)};
\ifthenelse{\isodd{#4}}{\draw[decorate] (i) -> (h)}{};
\ifthenelse{\isodd{#5}}{\draw[decorate] (j) -> (h)}{};
\ifthenelse{\isodd{#6}}{\draw[decorate] (j) -> (l)}{};
\end{tikzpicture}}

\newcommand{\od}{\text{od}}
\newcommand{\id}{\text{id}}
\newcommand{\odp}{\text{od\_av}}
\newcommand{\odd}{\text{odd}}

\begin{document}

\maketitle
\begin{abstract}
Stochastic actor-oriented models (SAOM) are a broadly applied modelling
framework for analysing network dynamics using network panel data.
They have been extended to address co-evolution of multiple networks
as well as networks and behaviour.
This paper extends the SAOM to the analysis of multiple network panels
through a random coefficient multilevel model, estimated with a Bayesian approach.
This is illustrated by a study of the dynamic interdependence of
friendship and minor delinquency, represented by the combination of a
one-mode and a two-mode network, using a sample of 81 school classes
in the first year of secondary school.
\end{abstract}



\section{Introduction}

Social network research deals with analysing the dependencies among people
or other social units, dependencies induced by the relational ties that bind them
together  \citep{WassermanFaust94,brandes2013network,robins2015doing}.
These dependencies can best be studied in a dynamic approach,
where the existence of a given configuration of ties leads to the
creation, or supports the maintenance, of other ties.
While many of the endogenous network dependencies, like
triadic closure and balance, are of interest in their own right,
there is a growing interest in the dynamic interdependence
of networks with other structures,
such as actor variables \citep{VeenstraEtAl2013},
other networks for the same actor set \citep{HuitsingSvDV2014,ElmerBodaStadtfeld2017},
or two-mode networks \citep{LomiStadtfeld2014}.

Dynamic network data can be of various kinds.
A frequently followed design is the collection of network panel data,
i.e., the observation of all relational ties (in one or more networks)
and other relevant variables,
within a given group of social actors (such as individuals, firms, countries, etc.),
at two or more moments in time, the `panel waves'.
For modelling panel data for a single network, represented by a digraph,
the Stochastic Actor-oriented Model (`SAOM') was proposed by \citet{Snijders01}.
This was extended to a joint model for changing
actor variables (vertex attributes) and tie-variables by \citet{SteglichEA10}
and to a model for the interdependent dynamics of multiple
networks, potentially combinations of one-mode and two-mode networks,
by \citet{SLT2013}.
These joint dynamic models can be combined under the heading
of `co-evolution', as summarized in \citet{Snijders2017}.

Collecting longitudinal network data is very time-intensive and demands
great care, but data sets of longitudinal networks in many
`parallel' groups are becoming increasingly common;
examples (among many others) are
the study `Networks and actor attributes in early adolescence' executed
by Chris Baerveldt and Andrea Knecht which will be used in this paper
\citep{Knecht2006,KnechtEA2010},
and CILS4EU \citep{CILS4EU}.

While the SAOM has proved useful in analysing networks in single groups,
the methodology has been limited in studying the extent to
which network dynamics generalise to different contexts and
what might differ systematically across groups of actors.
The investigation of heterogeneity across groups
more generally, in the way multilevel models have proven
useful \citep[e.g.,][]{Goldstein2011,SnijdersBosker12},
has not been possible.
However, to find scientific regularities it is more attractive to
study multiple groups that may be regarded as a sample from a population
and to generalize to populations of networks
\citep{SnijdersBaerveldt03,EntwisleEtAl07}.
For the Exponential Random Graph Model a multilevel methodology
was proposed by \citet{SlaughterKoehly2016} \citep[see also][]{schweinberger2017}.

This paper proposes a multilevel extension of the SAOM for data sets
composed of disjoint groups of actors, for which only networks
within each group are considered. The actors are nested within the groups.
Since ties combine pairs of actors, the combined structure of actors
and ties cannot be regarded as being nested.
This extension employs random coefficients like the multilevel
models mentioned above and draws on the likelihood-based estimation
frameworks of \citet{KoskinenSnijders07} and \citet{SnijdersEA10a}.
It also permits the investigation of observable group-level variables,
such as compositional and contextual factors, like in standard multilevel modelling.
Our example is a co-evolution of friendship networks and
delinquent behaviour represented by two-mode networks,
therefore the elaboration focuses on the
co-evolution model of \citet{SLT2013}.

Combinations of networks are occasionally refered to as `multilayer networks' \citep{KivelaEtAl2014,MagnaniWasserman2017} or `multilevel networks' \citep{Snijders16}, but in this paper we use the term `multilevel networks' to express the link to the random coefficient multilevel models in the sense mentioned above.





\section{Friendship and delinquency}

As the motivating example, we consider the dynamic relation between
friendship and delinquent behaviour, using the
study `Networks and actor attributes in early adolescence'.
The data set was collected by Andrea Knecht,
supervised by Chris Baerveldt \citep{Knecht2006}.
The data was collected
in 126 first-grade classrooms in 14 secondary
schools in The Netherlands in 2003-2004, using written questionnaires.
The entire data set contains four waves with about three months in between.

We focus on the friendship network and on the four questions
about delinquency:
stealing, vandalism, graffiti, and fighting, for each of which
self-reported frequencies were given with five categories.
Written self-reports provide reliable measurements of delinquency for
adolescents \citep{KollischOberwittler2004}.
The dynamic relation between a network such as friendship and a changing
actor variable such as the tendency to commit delinquent behaviour
has two sides: \emph{selection}, changes of friendships
as a function of the delinquent behaviour of the two individuals
concerned; and \emph{influence}, changes in delinquent behaviour
of an actor as a function of the network position of this actor
and the delinquent behaviour of the others, especially those to whom
this actor has a friendship tie.
A methodology to distinguish between selection and influence,
using network and behaviour panel data, based on the SAOM,
was proposed by \citet{SteglichEA10}.
The conclusions are not causal in the counterfactual sense,
as demonstrated by \citet{ShaliziThomas2011}, but in a temporal
sense: does a change in behaviour follow on some network configuration
(`influence'), or does a change in friendship follow on
a behaviour configuration (`selection'). A further discussion of
causality in network-behaviour systems was given by \citet{LSST11}.

The association between friendship and the tendency to delinquent
behaviour was studied by \citet{KnechtEA2010}. This publication used the
data set mentioned above, constructing an actor variable
representing delinquent behaviour as a sum score of the four
items for the frequencies of stealing, vandalism, graffiti, and fighting.
It used the two-step multilevel method of \citet{SnijdersBaerveldt03},
in which first the SAOM is estimated for each classroom separately,
after which the results for the classrooms are combined.
Since most of the classrooms were too small for the satisfactory
application of this --- rather complicated --- model, only 21 classrooms
could be used.

In the current paper we present an extension of this study,
replacing the simplistic two-step multilevel approach by an integrated
random coefficient approach, which does not depend on the condition
of a convergent estimation algorithm for each classroom separately and
therefore can use a much larger part of the data set.
Furthermore, we replace the model where delinquent behaviour is represented
by an actor variable with a model representing
the four delinquency items by a two-mode network.
This allows a more detailed study of social influence.
The actors are supposed to be influenced by their friends,
which are those they mention as a friend
(friendship ties from the actor to the friends).
In the former study, the tendency toward delinquent behaviour
was regarded as a one-dimensional trait, measured by the
sum score of the four delinquent items; social influence
was represented by the effect of the average of this trait
over the actor's friends. The current study considers this
together with another type of influence: the effect of
the friends' behaviour for some specific delinquent behaviour
on the same behaviour of the actor.

\section{Multilevel stochastic actor-oriented model}

The stochastic actor-oriented model \citep{Snijders2017}
is a family of longitudinal network models for network panel data.
While networks are only observed
at discrete time points, the model assumes that the
networks evolve in continuous time.
This is necessary for representing the feedback between the
tie variables that can occur in the time elapsing between
the observation moments.
Some history of continuous-time models for social network
panel data is presented in \citet{Snijders01}.
Continuous-time models for discrete-time panel data are well known
\citep[e.g.,][]{Bergstrom1988,HamerleSingerNagl1993,Singer1996}.
Their use for network panel data in sociology
is argued also by \citet{Block2018}.

\subsection{Data structure}

We assume that we have panel network data for $G$ independent groups.
The groups are a collection of mutually exclusive fixed sets
of nodes $\mathcal{N}_1,\ldots, \mathcal{N}_G$, with time-dependent
one-mode networks for each of them. In our example, these nodes represent individuals
and the network represents the friendships among them.
We assume that there may
only be network ties between nodes in the same node sets,
and at any point in time $t$, the network in group $\mathcal{N}_g$ is
represented by a binary adjacency matrix
$X^{[g]}(t)= (X_{ij}^{[g]}(t) )_{(i,j) \in \mathcal{N}_g \times  \mathcal{N}_g }$,
where $X_{ij}^{[g]}(t) =1$ if there is a tie from $i$ to $j$ at time $t$, and zero otherwise.
Self-ties are excluded.
In addition, we have two-mode networks with a common second-mode node set
$\mathcal{H}$, which here is the set of the $H=4$ delinquency behaviours.
The delinquency behaviours are dichotomized, and $Z_{ih}^{[g]}(t)$ indicates whether individual $i$ in group $g$ engages in behaviour $h$ at time $t$. These two-mode tie variables are collected in a matrix $Z^{[g]}(t)$.
Jointly we denote the one-mode and two-mode network by
$Y^{[g]}(t)=( X^{[g]}(t) , Z^{[g]}(t) )$, for $g=1,\ldots, G$.
The supports of $X^{[g]}$ and $Z^{[g]}$
are denoted $\mathcal{X}_g$ and $\mathcal{Z}_g$,
respectively, with joint support $\mathcal{Y}_g=\mathcal{X}_g \times \mathcal{Z}_g$.

For the data, we assume that $Y^{[g]}(t)$ is observed at discrete points in time,
$t_0,t_1,\ldots,t_M$, where $M$ can be as small as~2.
The inferential target is to model how $Y^{[g]}(t_{m-1})$
changed into $Y^{[g]}(t_{m})$ for $m = 1, \ldots, M-1$.

\subsection{Model specification for a single group}

The model for the SAOM in a single group can be described without
the notational dependence on the group membership,
as one-mode ties are only defined within groups.
Therefore we drop the superscript ${[g]}$.
The process is \emph{actor-oriented} in the sense that transitions in the
process are modelled as choices by actors $i \in \mathcal{N}$
to change outgoing tie variables $X_{ij}$ or $Z_{ih}$.
It is assumed that $Y(t)$, $t_1 \leq t \leq t_M$, given the available covariates,
is a Markov process in continuous time.
We present the SAOM for the case of co-evolution of a one-mode and a two-mode
network; this can be generalized to more networks
and to co-evolution with behavioural variables, see \citet{Snijders2017}.

At any moment $t$ in continuous time, at most one actor $i$
may make a change in at most one tie variable $X_{ij}$ or $Z_{ih}$;
this can be creation of the tie (0 to 1) or termination (1 to 0).
This restriction was proposed already by \citet{HoLe77a}, and it implies
that the dynamic model is decomposed in the smallest possible changes;
these changes are called \emph{mini-steps}.
Basic ingredients of the model are \emph{rate functions}
$\lambda_i^X(\theta,y)$ and $\lambda_i^Z(\theta,y)$ which indicate
the rates at which actor $i$ gets an opportunity, respectively,
to change some one-mode tie $X_{ij}\ (j \in \mathcal{N}, j \neq i)$
or  to change some two-mode tie $Z_{ij}\ (j \in \mathcal{H})$;
and \emph{evaluation functions} $f_i^X(\theta,y)$ and $f_i^Z(\theta,y)$
indicating the value, as it were, that actor $i$ attaches
to state $y$ of the combined networks when making, respectively,
a change in network $X$ or in network $Z$.
The rate functions define the expected frequency of the mini-steps
and the evaluation functions define the probability distribution of their results.
For simple models the number of opportunities has a Poisson distribution.
Since the choice situations with respect to the one-mode network (friendship)
and the two-mode network (delinquency behaviour) are different,
different considerations for the actors may apply, and the
evaluation functions $f_i^X(\theta,y)$ and $f_i^Z(\theta,y)$
will not be the same.

By the properties of the exponential distribution, the time until the first
opportunity for change of any kind by any actor is exponentially distributed with rate
\[
\lambda^+_+(\theta,y) \,=\, \sum_{i \in \mathcal{N}} \big( \lambda^X_i(\theta,y)
        \,+\, \lambda^Z_i(\theta,y) \big) \ ,
\]
and the probability that actor $i \in \mathcal{N}$ is selected for changing
a tie variable in $V \in \{X, Z\}$  is
\[
        \frac{\lambda^V_i(\theta,y)}{\lambda^+_+(\theta,y)} \ .
\]

Given that $i$ is selected for making a change in network $V$,
the option set consists of all outgoing tie variables in network $V$,
together with the option `no change'.
The set of outcomes reachable in a mini-step  by actor $i$ in network
$V$ is denoted $\mathcal{A}^V_i(y)$, with
\[
 \mathcal{A}^X_i(x,z) \, \subseteq \,
    \{ (x^{\prime}, z) \in \mathcal{Y}: ||x-x^{\prime} || \leq 1,
                x^{\prime}_{kj}=x_{kj},\, \forall j \text{ and } \forall k \neq i\}
\]
and
\[
 \mathcal{A}^Z_i(x,z) \, \subseteq \,
    \{ (x, z^{\prime}) \in \mathcal{Y}: ||z-z^{\prime} || \leq 1,
                z^{\prime}_{kh}=z_{kh},\, \forall h \text{ and } \forall k \neq i\} \ .
\]
Here $||B-C||$ denotes the Hamming distance between
adjacency matrices $B$ and $C$.
Usually the subset ``$\subseteq$'' will be implemented as equality ``$=$'',
but the subset symbol is used because there could be constraints on the state space,
such as in the case of changing composition or absorbing states.

Conditionally on $y$, and on $i$ being selected to make a change in network $V$,
the probability that the outcome of the choice is $y^{\prime}$ is
\begin{equation}
 p^V_i(\theta, y, y^{\prime}) \,=\,
 \frac{ \exp\big(f^V_i(\theta,y^{\prime})\big) }
  { \sum_{\tilde y \in \mathcal{A}^V_i(y)}  \exp\big(f^V_i(\theta,\tilde y)\big)} \,=\,
 \frac{ \exp\big(f^V_i(\theta,y^{\prime}) - f^V_i(\theta,y)\big) }
  { \sum_{\tilde y \in \mathcal{A}^V_i(y)} 
   \exp\big(f^V_i(\theta,\tilde y)  - f^V_i(\theta,y) \big)}
        \label{choiceprob}
\end{equation}
if $ y^{\prime} \in \mathcal{A}^{V}_{i}(y)$,
and 0 if $y^{\prime} \not\in \mathcal{A}^V_i(y)$.
Note that since $y \in \mathcal{A}^{V}_{i}(y)$,
the probability of no change, i.e., $y^{\prime}=y$, is positive.

\subsubsection{Interpretation of process}
\label{S_interp}

For notational convenience, we further use the symbol $y$
instead of $y^{\prime}$ in the role of outcome of the mini-step.
Typically, the evaluation functions $f^V_i(\theta,y)$
are modelled as weighted functions of statistics calculated on $y$,
\[
    f^V_i(\theta,y) \,=\, \sum_k \theta^V_k \, s_{ki}^V(y) \ .
\]
The statistics $s_{ki}^V(y)$ are briefly called \emph{effects}, and will be
functions pertaining to actor $i$ and the network neigbourhood of $i$.
Usual effects $s_{ki}^V(x,z)$ are counts of subgraphs (configurations)
that include ties originating with actor $i$.
Since no information is available on the timing of the mini-steps,
the focus of modeling is on the evaluation functions and not on the
rate functions \citep[an exception is the diffusion model of][]{Greenan15}.
Often the rate functions $\lambda^X_i(\theta,y)$
and $\lambda^Z_i(\theta,y)$ are chosen to be constant between observation
moments, and that is what will be assumed further on.
If the evaluation function $f^X_i\big(\theta,(x,z)\big)$ does not depend
on $z$ and $f^Z_i\big(\theta,(x,z)\big)$ does not depend on $x$, the dynamics
of the one-mode and two-mode networks are independent.
In our example the interest is in the interdependence between friendship
and delinquent behaviour, which is reflected by statistics that depend
on both networks jointly.

The model can be interpreted as a sequential discrete-choice model
where actors make choices about their outgoing ties,
using random utilities \citep{Maddala83}, under the restriction that
they can change no more than one outgoing tie variable.
From that perspective the model can be interpreted as a process whereby actors
chose to change their network ties or their behaviour to what they deem most preferable,
allowing for a random element in their decisions.
The model does not strictly require this interpretation and
\citet{Snijders2017} treats a wide variety of different model specifications,
including differential treatments of creating and terminating ties,
more elaborate specifications of the rate functions, and options
for non-directed networks.

Of particular importance are cross-network effects $s_{ki}^X(x,z)$ and
$s_{ki}^Z(x,z)$ depending on $x$ as well as $z$,
reflecting the mutual dependence between the one-mode and the two-mode network.
In our application, where the networks are friendship and delinquent behaviours,
the following cross-network effects are used.
As mnemonic indicators, we use `o' for outgoing friendship ties,
'i' for incoming friendship ties, and 'd' for ties in the delinquency network.
The subgraphs used are illustrated in the pictograms, where nodes of the first mode
are denoted by circles, nodes of the second mode by squares, one-mode ties
by straight arrows, and two-mode ties by curly arrows.
The superscript $V$ indicates that the effect applies to $V=X$ as well as $V=Z$.
Note that effects $s_{ki}^V$ refers to actors $i$, who consider changing some
outgoing tie in network $V$.
In the pictograms, the parts with a tie $i \rightarrow j$ have the role of
dependent variables for friendship, and the parts with a tie
$i \rightsquigarrow h$ have the role of dependent variables for delinquency.

\begin{enumerate}
  \item \od: the product of the number of outgoing friendships and the number
       of delinquent behaviours of $i$,\label{outActIntn}
   \begin{align*}
        s_{\od,i}^V(x,z) \,=\, \sum_{j} x_{ij} \sum_{h} z_{ih} \ .
         &&&&  \raisebox{-0.5em}{  \motif{-4}{0}{1}{1}{0}{0}{1} }
  \end{align*}
  \item \id: the product of the number of incoming friendships and the number
       of delinquent behaviours (note the exchange of $i$ and $j$),
\label{inActIntn} 
  \begin{align*}
        s_{\id,i}^X(x,z) \,=\, \sum_{j} x_{ij} \sum_{h} z_{jh} \ ,
         &&&&   \raisebox{-0.5em}{ \motif{-4}{0}{0}{1}{0}{0}{0} } \\
        s_{\id,i}^Z(x,z) \,=\, \sum_{j} x_{ji} \sum_{h} z_{ih} \ .
         &&&& \raisebox{-0.5em}{ \motif{-4}{0}{0}{1}{0}{0}{1} }
  \end{align*}
  \item \odd: a mixed triadic effect: the number of friendships of $i$
       weighted by the number of delinquent  behaviours $i$ and $j$ have in  common,
\label{tofrom}
   \begin{align*}
        s_{\odd,i}^V(x,z) \,=\, \sum_{j,h} x_{ij} \, z_{ih} \, z_{jh} \ .
        &&&& \raisebox{-2em}{ \motif{2}{3}{1}{1}{1}{0}{1}  }
  \end{align*}
  \item \odp: a mixed four-node effect that is not a subgraph count:
        the total number of delinquent behaviours reported by $i$
        multiplied by the average number of delinquent behaviours, centered,
        reported by all $i$'s friends,\label{outOutAvIntn}
  \begin{align*}
        s_{\odp,i}^Z(x,z) \,=\, \sum_{h} z_{ih} \,
   & \frac{\sum_j x_{ij} \big\{ \big( \sum_{\ell} z_{j\ell} \big) - \bar z \big\} }
                    {\sum_j x_{ij} } \ , \\
        &  \hfill \raisebox{0em}{   \motif{-4}{0}{1}{1}{0}{1}{1} } \nonumber
  \end{align*}
    where $\bar z $ is the average observed outdegree for $Z$ in the group.
Here 0/0 is defined as 0.
\end{enumerate}
Effect `\odp' is used only for explaining the dynamics of the $Z$ network,
the other three are used for explaining the dynamics
of both networks. 
Brief interpretations of these effects, for positive parameter values,
are the following.

\noindent
For explaining the friendship dynamics (`selection'):
\begin{itemize}
  \item[(\ref{outActIntn})] The `\od'\ effect indicates that those who engage in
       more delinquent behaviours will be more active in nominating friends.
  \item[(\ref{inActIntn})] The `\id'\ effect indicates that those who engage in
       more delinquent behaviours will be more popular as friends.
  \item[(\ref{tofrom})] The `\odd'\ effect indicates that actors will tend to be friends
      with those who engage in the same delinquent behaviours.
\end{itemize}
And for explaining the delinquency dynamics (`influence'):
\begin{itemize}
  \item[(\ref{outActIntn})] The `\od'\ effect indicates that those who nominate
      more friends will tend to engage in more delinquent behaviours.
  \item[(\ref{inActIntn})] The `\id'\ effect indicates that those who are
       more popular as friends will tend to engage in more delinquent behaviours.
  \item[(\ref{tofrom})] The `\odd'\ effect indicates that actors will tend to
      engage in the same delinquent behaviours as their friends.
  \item[(\ref{outOutAvIntn})] The `\odp'\ effect indicates that those whose friends
       on average are more delinquent will also themselves tend to engage
       in more delinquent behaviours.
\end{itemize}
The last two effects (`\odd'\ and `\odp') are
the most clear expressions of the idea of social influence, both implying that
the probability distribution of changes in delinquent behaviour of the actor is
a function of the delinquent behaviour of the actor's friends.
Effect `\odd'\ is social influence operating for specific acts of delinquent
behaviour, while `\odp' is a generalized influence at the level
of the sum scores of delinquency.

\subsection{Data augmentation}

The SAOM with rates $(\lambda_i^V(\theta,y))$
and one-step jump probabilities $(p^V_i(\theta,y, y^{\prime}))$
defines a discrete Markov chain in continuous time
with intensity matrix defined for $y \neq y^{\prime}$,
and $V \in \{X, Z\}$, by
\begin{equation}
  q(y, y^{\prime}) = \left\{
    \begin{array}{ll}
     \lambda^V_i(\theta, y)\,  p^V_i(\theta,y, y^{\prime})    &
                    \text{ if } y^{\prime} \in  \mathcal{A}^V_i(y) \\
     0 & \text{ otherwise.}
    \end{array}
     \right.                                                     \label{q}
\end{equation}
The process can be defined as a marked point process. Only in trivial cases,
such as the random walk on a $|\mathcal{Y}|$-cube \citep{aldous1983},
is Bayesian inference for such models tractable \citep{KoskinenSnijders07}.
For two waves of observations $y(t_m)$ and $y(t_{m+1})$, the likelihood is
a $|\mathcal{Y}|$ times $|\mathcal{Y}|$ matrix
\[
P^{T} = e^{TQ}{\text{,}}
\]
for $T=t_{m+1}-t_{m}$, which is huge. The model is doubly intractable
given that both the likelihood and the posterior involve intractable
normalising constants.
Index the mini-steps by $r = 1, \ldots, R$ (where $R$ is random), and
denote the results of the mini-steps by $v^r = (i^r, V^r, y^r)$
and the holding times by $(s^r)$.
Koskinen and Snijders (2007) propose to augment data by
performing joint inference over the model parameters $\theta$ as well as the
unobserved sequences $(v^r)$ and $(s^r)$.
The sequence $(i^r, V^r, y^r)$ must be such that
if $y^r$ differs from $y^{r-1}$ it is only in variable $V^r$
and row $i^r$ of the adjacency matrix.
The augmented data likelihood, conditional on $y^0=y(t_m)$,
for a sequence of holding times $(s^r)$
and results of mini-steps $v = (v^r) = \big((i^r, V^r, y^r)\big)$, is given by
\begin{align*}
p^{\ast}_{\mathrm{AUG}}\big((v^r),(s^r) \,|\, y^0,\theta\big) \,=\,&
\exp\Bigg\{ -\sum_{r=1}^{R} s^r \lambda_{+}^+(\theta,y^{r-1}) \Bigg\}\\
 & \times \prod_{r=1}^{R}\lambda_{i_r}^{V_r}(\theta,y^{r-1})\,
                p_{i_r}^{V_r}(\theta, y^{r-1}, y^{r}) \ .
\end{align*}
It is more efficient to work with the marginal model
$p_{\mathrm{AUG}}\big( v \,|\, y(t_{m})\big)$ which is
$p^{\ast}_{\mathrm{AUG}}\big( v , s \,|\, y(t_{m})\big)$
marginalised over holding times $s$.
In the sequel we will assume constant rates
$\lambda_i^V=\lambda^V$ for both networks $V=X,Z$,
in which case the augmented likelihood is
\begin{align} \label{augm}
p_{\mathrm{AUG}}\big((v^r) \,|\, y^0,\theta\big) & \,=\,
\exp\big( - \lambda_{+}^+ (t_{m+1}-t_m)\big) \\
 \times & \frac{\big( \lambda_{+}^+ (t_{m+1}-t_m)\big)^R }{R!}
 \prod_{r=1}^{R} \Big( \frac{\lambda^{V_r}}{\lambda^X + \lambda^Z}\Big)\,
                p_{i_r}^{V_r}(\theta, y^{r-1}, y^{r}) \ ; \nonumber
\end{align}
see \citet{SnijdersEA10a} where also an approximation
for non-constant rates is given.

The Markov assumption implies that the likelihood for
a sequence of augmented data  $v = \big(v(t_1), \ldots, v(t_M)\big)$,
given observation $y = \big(y(t_{0}), y(t_{1}), \ldots, y(t_{m})\big)$ is
\begin{equation}  \label{AUG}
  p_{\mathrm{AUG}}(v \,|\, y, \theta ) \,=\,
        \prod_{m=1}^{M-1} p_{\mathrm{AUG}}\big(v(t_{m+1}) \,|\, y(t_{m}),\theta \big) \ .
\end{equation}

The model $p_{\mathrm{AUG}}\big( v  | y(t_{m-1}), \theta \big)$
when marginalised over all paths $v$ that start in $y^0 = y(t_{m-1})$ and end in
$y^R=y(t_{m})$ is the data likelihood per wave
\[
p_{\mathrm{SAOM}}\big(y(t_{m}),\theta \big) \  
\]
with the obvious extension for multiple waves.

\section{Hierarchical model}

We assume that each group $g$ follows the same specification,
i.e., has the same expressions for the rate and evaluation functions, although
the number of actors $n_g = | \mathcal{N}_g|$ may be different.
Each group $g$ has associated with it a group-specific
parameter $\theta^{[g]}$. Heterogeneity across groups typically takes the
form of contextual and compositional effects.

While comparing structure across networks is a natural thing to do and has attracted
some attention \citep[e.g.,][]{faust2002}, it is clear that comparing structure across
different-sized networks is non-trivial \citep{anderson1999}.
One key problem is the way
the average degree scales with network size, something that has
been studied for cross-sectional networks
\citep{ErdosRenyi1960,krivitsky2011,shalizi2013}.
We assume that the variation in group sizes $n_g$ as well as
in average degrees is limited.
Based on a combination of \citet{krivitsky2011} and
\citet[][p. 243]{Snijders05} we suggest that including
an effect of $\log(n_g) \sum_j x_{ij}$ will make the other parameters
comparable, and that for this effect a parameter of $-1/2$
would be expected if none of the other parameters
reflects differential group sizes.

Components of $\theta^{[g]}$ that are variable across $g$
are similar to random slopes in regular multilevel modeling
\citep{Goldstein2011,SnijdersBosker12}.
The question of whether to allow all group-level parameters to vary across groups
needs to be guided by specific case considerations as well as
computational aspects just as in multilevel models in general.
We partition the parameter vector  $\theta^{[g]}$ for group $g$
into subvectors $\gamma^{[g]}$, of dimension $p_1$,
containing the variable parameters,
and $\eta$, of dimension $p_2$, containing the constant parameters.
We write the group-wise parameters as the partitioned vector
\begin{equation*}
\theta^{[g]}=\left(
\begin{array}{c}
\gamma^{[g]} \\
\eta
\end{array}
\right) .
\end{equation*}
\\
When $p_1=0$ we have the so-called multi-group model \citep[Section 11.2]{SienaManual21}.
In classical multilevel modeling it is usual to apply models with
only a few random slopes. However, it seems that Bayesian estimation
allows entertaining models with more random slopes \citep{EagerRoy2017}.
For group-level covariates, such as interventions or indicators
of group composition, it is natural that their effects are fixed.

We draw on standard hierarchical modelling approaches and assume that the
group-level parameters have a multivariate normal distribution
$\gamma^{[g]} \stackrel{iid}{\thicksim} N_{p_1}(\mu,\Sigma)$.
We assume that $(\mu,\Sigma)$ and $\eta$ are a priori independent
with priors
$(\mu,\Sigma) \thicksim \pi(\mu,\Sigma \,|\, \Gamma)$ and
$\eta \thicksim \pi(\eta \,|\, \mu_{0,\eta},\Sigma_{0,\eta})$.

An exception to this should be made for the rate parameters $\lambda$,
which are necessarily positive.
They reflect particular circumstances of groups
and issues of study design, and will always be included
among the variable parameters $\gamma^{[g]}$. The multivariate
normal distribution is assumed to be truncated
to positive values for these parameters. The values of $\mu$ and $\Sigma$
will in practice be such that the non-truncated distribution has
an extremely small probability for negative rate parameters.
An alternative is to employ a transformed normal or a
Gamma distribution, which is conjugate for the Poisson
counts \citep{KoskinenSnijders07}.
However, the multivariate normal gives a simple unified treatment
for all varying parameters.

With this hierarchical specification, denoting the
multivariate normal density by $\phi$,
the joint probability density function for data $y^{[1]}, \ldots, y^{[G]}$,
parameters $\gamma^{[1]}, \ldots, \gamma^{[G]}$,
and $\mu, \Sigma, \eta$ is given by
\begin{equation}
\pi\big( \mu,\Sigma \mid \Gamma \big) \, \pi(\eta|\mu_{0,\eta},\Sigma_{0,\eta})
\prod_{g=1}^G \phi(\gamma^{[g]} \mid \mu, \Sigma) \,
        p_{\rm SAOM}(y^{[g]} \mid \gamma^{[g]}, \eta) \ .
\end{equation}

\section{Prior specifications}
We present the inference scheme for a specific choice of priors.
Other prior specifications may be considered (see Appendix B)
but the MCMC scheme largely remains unchanged.

\subsection{Varying parameters: conjugate prior}

For multivariate normal distributions with unknown
expected value $\mu$ and covariance matrix $\Sigma$, the conjugate prior
distribution is the inverse Wishart distribution for $\Sigma$,
and conditional on $\Sigma$ for $\mu$ a multivariate normal distribution:
\begin{itemize}
\item $\Sigma \sim
    \mathrm{InvWishart}_{p}(\Lambda_0,\nu_0)$, and
    conditionally on $\Sigma$
\item $\mu \mid \Sigma \sim N_p(\mu_0, \Sigma/\kappa_0)$ \ .
\end{itemize}
This is treated, e.g., in \citet{BDA3}, Section 3.6, and
\citet{HaganForster2004}, Chapter 14.
Thus, the hyper-parameters of the prior are $\Lambda_0, \nu_0, \kappa_0$.
The expected value for the inverse Wishart($\Lambda, \nu$) distribution is
\[
\E\big\{\Sigma\big\} = \frac{1}{\nu-p-1}\, \Lambda
\]
provided $\nu > p+1$,
and the mode is $(\nu+p+1)^{-1}\Lambda$ \citep{HaganForster2004}.
Thus, the central tendency of the inverse Wishart($\Lambda, \nu$)
distribution may be taken to be about $\nu^{-1}\Lambda$.
Parameter $\Lambda$ is on the scale of the sum of squares of a
sample of size $\nu$ from a distribution with variance-covariance
matrix $\Sigma$.
The number of degrees of freedom $\nu_0$ can be regarded as the
effective sample size that has led to the prior information.
The value of $\kappa_0$ can be interpreted as the proportionality
between $\Sigma$, the uncertainty about the groupwise parameters
$\gamma^{[g]}$ given the average population value $\mu$,
and the prior uncertainty about $\mu$.
Having the same proportionality of this kind for all parameters is
rather restrictive, but as a first approach we prefer to use a conjugate
prior which leads to relatively simple procedures
for this already complicated model.

\subsection{Constant parameters}

For most components of the group-constant parameter $\eta$ we assume
an improper prior with constant density $\pi(\eta) \propto c$.
This is justified because for the estimation
of $\eta$ the information from all groups is combined,
leading for $\eta$ to a quite weak dependence on the prior.
However, for effects of group-level covariates the situation is different, and
for those components of $\eta$ a multivariate normal prior distribution will be assumed.

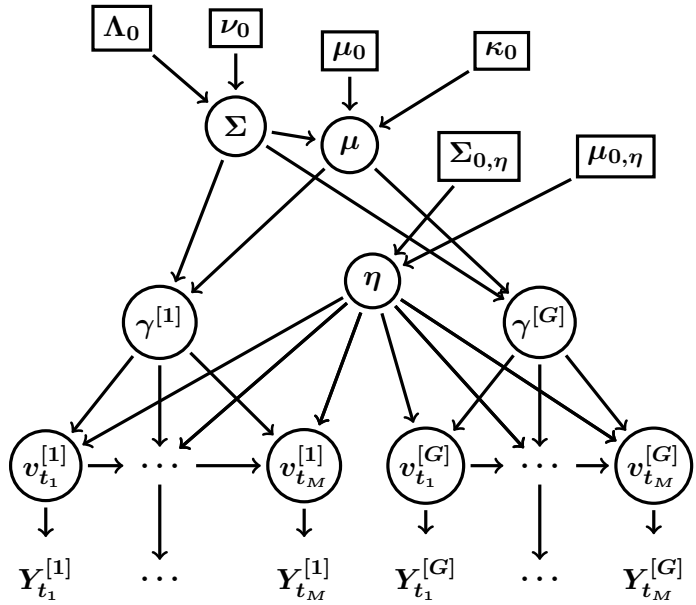
\begin{figure}
	\begin{center}
	\begin{tikzpicture}
	\begin{scope}
	\boldmath
	\tikzstyle{every path}=[very thick, -stealth', shorten <=2pt,
	shorten >=2pt, -]

	\node[shape=rectangle,draw] (A) at (1,5.3) { $\Lambda_0$ };
	\node[shape=rectangle,draw] (B) at (2.5,5.3) { $\nu_0$ };
	\node[shape=rectangle,draw] (C) at (4,5) {  $ \mu_0 $  };
	\node[shape=rectangle,draw] (D) at (6,5) {  $ \kappa_0 $ };

	\node[shape=rectangle,draw] (Apa) at (5.7,3.6) { $\Sigma_{0,\eta}$ };
	\node[shape=rectangle,draw] (Bpa) at (7.5,3.6) { $\mu_{0,\eta}$ };

	\node[shape=circle,draw] (A2) at (2.5,4) {  $\Sigma$    };
	\node[shape=circle,draw] (B2) at (4,3.75) {  $\mu$  };

	\node[shape=circle,draw] (AApa) at (4.3,1.95) {  $\eta$  };

	\node[shape=circle,draw,inner sep=2pt]  (C2) at (1.5,1.4) {  $\gamma^{[1]}$   };

	\node[shape=circle,draw,inner sep=1pt](E2) at (6.5,1.4) {  $\gamma^{[G]}$   };

	\node[shape=circle,draw,inner sep=1pt]  (CA2) at (0,-0.5) {  $v^{[1]}_{t_1}$   };

	\node(DA2) at (1.5,-0.5) { $\ldots$ } ;

	\node[shape=circle,draw,inner sep=1pt] at (3.4,-0.5)(EA2) { $v^{[1]}_{t_M}$     };

	\node[shape=circle,draw,inner sep=1pt] (CB2) at (5,-0.5) {  $v^{[G]}_{t_1}$   };

	\node(DB2) at (6.5,-0.5) { $\ldots$ } ;

	\node[shape=circle,draw,inner sep=1pt](EB2) at (8,-0.5) { $v^{[G]}_{t_M}$     };

	\node[shape=circle,inner sep=0pt]  (CA3) at (0,-2.0) {  $Y^{[1]}_{t_1}$   };

	\node(DA3) at (1.5,-2.0) { $\ldots$ } ;

	\node[shape=circle,inner sep=0pt] at (3.4,-2.0)(EA3) { $Y^{[1]}_{t_M}$     };

	\node[shape=circle,inner sep=0pt]  (CB3) at (5,-2.0) {  $Y^{[G]}_{t_1}$   };

	\node(DB3) at (6.5,-2.0) { $\ldots$ } ;

	\node[shape=circle,inner sep=0pt](EB3) at (8,-2.0) { $Y^{[G]}_{t_M}$     };

	\draw[very thick, -stealth', shorten <=2pt, shorten >=2pt, ->]  (A) -> (A2);
	\draw[very thick, -stealth', shorten <=2pt, shorten >=2pt, ->]  (B) -> (A2);

	\draw[very thick, -stealth', shorten <=2pt, shorten >=2pt, ->]  (C) -> (B2);
	\draw[very thick, -stealth', shorten <=2pt, shorten >=2pt, ->]  (D) -> (B2);

	\draw[very thick, -stealth', shorten <=2pt, shorten >=2pt, ->]  (A2) -> (B2);

	\draw[very thick, -stealth', shorten <=2pt, shorten >=2pt, ->]  (A2) -> (C2);

	\draw[very thick, -stealth', shorten <=2pt, shorten >=2pt, ->]  (B2) -> (C2);



	\draw[very thick, -stealth', shorten <=2pt, shorten >=2pt, ->]  (A2) -> (E2);

	\draw[very thick, -stealth', shorten <=2pt, shorten >=2pt, ->]  (B2) -> (E2);

	\draw[very thick, -stealth', shorten <=2pt, shorten >=2pt, ->]  (C2) -> (CA2);

	\draw[very thick, -stealth', shorten <=2pt, shorten >=2pt, ->]  (C2) -> (DA2);

	\draw[very thick, -stealth', shorten <=2pt, shorten >=2pt, ->]  (C2) -> (EA2);

	\draw[very thick, -stealth', shorten <=2pt, shorten >=2pt, ->]  (E2) -> (CB2);

	\draw[very thick, -stealth', shorten <=2pt, shorten >=2pt, ->]  (E2) -> (DB2);

	\draw[very thick, -stealth', shorten <=2pt, shorten >=2pt, ->]  (E2) -> (EB2);

	\draw[very thick, -stealth', shorten <=2pt, shorten >=2pt, ->]  (CA2) -> (DA2);

	\draw[very thick, -stealth', shorten <=2pt, shorten >=2pt, ->]  (DA2) -> (EA2);

	\draw[very thick, -stealth', shorten <=2pt, shorten >=2pt, ->]  (CB2) -> (DB2);

	\draw[very thick, -stealth', shorten <=2pt, shorten >=2pt, ->]  (DB2) -> (EB2);

	\draw[very thick, -stealth', shorten <=2pt, shorten >=2pt, ->]  (Apa) -> (AApa);

	\draw[very thick, -stealth', shorten <=2pt, shorten >=2pt, ->]  (Bpa) -> (AApa);

	\draw[very thick, -stealth', shorten <=2pt, shorten >=2pt, ->]  (AApa) -> (CA2);

	\draw[very thick, -stealth', shorten <=2pt, shorten >=2pt, ->]  (AApa) -> (DA2);

	\draw[very thick, -stealth', shorten <=2pt, shorten >=2pt, ->]  (AApa)-> (EA2);

	\draw[very thick, -stealth', shorten <=2pt, shorten >=2pt, ->]  (AApa)-> (CB2);

	\draw[very thick, -stealth', shorten <=2pt, shorten >=2pt, ->] (AApa) -> (DB2);

	\draw[very thick, -stealth', shorten <=2pt, shorten >=2pt, ->]  (AApa)-> (EB2);

	\draw[very thick, -stealth', shorten <=2pt, shorten >=2pt, ->]  (AApa)-> (DA2);

	\draw[very thick, -stealth', shorten <=2pt, shorten >=2pt, ->] (AApa)-> (EA2);

	\draw[very thick, -stealth', shorten <=2pt, shorten >=2pt, ->]  (AApa) -> (DB2);

	\draw[very thick, -stealth', shorten <=2pt, shorten >=2pt, ->] (AApa) -> (EB2);

	\draw[very thick, -stealth', shorten <=2pt, shorten >=2pt, ->] (CA2) -> (CA3);

	\draw[very thick, -stealth', shorten <=2pt, shorten >=2pt, ->] (DA2) -> (DA3);

	\draw[very thick, -stealth', shorten <=2pt, shorten >=2pt, ->] (EA2) -> (EA3);

	\draw[very thick, -stealth', shorten <=2pt, shorten >=2pt, ->] (CB2) -> (CB3);

	\draw[very thick, -stealth', shorten <=2pt, shorten >=2pt, ->] (DB2) -> (DB3);

	\draw[very thick, -stealth', shorten <=2pt, shorten >=2pt, ->] (EB2) -> (EB3);

	\end{scope}
	\end{tikzpicture}
\end{center}
	\caption{Dependence structure of hierarchical SAOM,
representing only the first and last groups $g=1,G$.}\label{fig:dag}
\end{figure}

\section{Estimation}
\label{S_estimation}

The dependence structure amongst all variables is given in Figure~\ref{fig:dag}.
Parameters can be estimated by an MCMC procedure,
sampling the random variables indicated by the circles in Figure~\ref{fig:dag},
going up in the figure.
The parameters in rectangular boxes are given hyperparameters.

\emph{\textbf{Mini-steps}}\\
For all groups $g$ independently,
sequences $v^{[g]}$ of outcomes of mini-steps $(i^r, V^r, y^r)$
are sampled by an extension of the Metropolis-Hastings
procedures of \citet{KoskinenSnijders07} and \citet{SnijdersEA10a}.
The extension consists of the insertion of the determination of $V^r$.
The target probability function is (\ref{AUG})
for given $y=y^{[g]}$ and $\theta = (\gamma^{[g]}, \eta)$.

\newpage
\emph{\textbf{Groupwise varying parameters}}\\
Groupwise varying parameters $\gamma^{[g]}$ are sampled
for given $v^{[g]}$ and $\eta, \mu, \Sigma$,
again for all groups $g$ independently,
by Metropolis Hastings steps with target density
\[
 \phi(\gamma^{[g]} \mid \mu, \Sigma) \,
       p_{\mathrm{AUG}}(v^{[g]} \,|\, \gamma^{[g]}, \eta )  \ .
\]
Here $\phi$ is the multivariate normal density and $p_{\mathrm{AUG}}$
was given in (\ref{AUG}).
A random walk proposal distribution is used, like in
\citet[][Ch. 5.4]{Schweinberger2007} and
\citet[][Section 4.4]{KoskinenSnijders07}.
The covariance  matrix for the proposals is $C^{[g]}$ as defined below
in the section on initial values,
scaled to obtain approximately  25 \% acceptance rates \citep{GRG96}.

\emph{\textbf{Constant group-level parameters}}\\

The constant parameter $\eta$ with prior density
$\pi(\eta \,|\, \mu_{0,\eta}, \Sigma_{0,\eta})$ can be sampled in two ways,
both using Metropolis Hastings steps analogous to the sampling of
the groupwise varying parameters.

The first way draws random walk proposals for $\eta$
with additive perturbations from the multivariate normal distribution
with mean 0 and covariance matrix  $C^{[0]}_\eta$
given below, scaled to obtain approximately  25 \% acceptance rates.
The target distribution is
\[
\prod_{g=1}^G \pi(\eta \,|\, \mu_{0,\eta}, \Sigma_{0,\eta})\,
         p_{\mathrm{AUG}}(v^{[g]} \,|\, y^{[g]}, \gamma^{[g]}, \eta ) \ .
\]

The second way draws random walk proposals for additive changes
in the entire vectors $\big(\gamma^{[g]}, \eta\big)$, excluding
the basic rate parameters. Now the perturbations
come from the multivariate normal distribution
with mean 0 and covariance matrix  $C^{[0]}$ given below, again
scaled to obtain approximately  25 \% acceptance rates.
The proposal is to add this perturbation identically to the vectors
$\big(\theta^{[g]}, \eta\big)$ for all $j$.
The target distribution is
\[
\prod_{g=1}^G \pi(\eta \,|\, \mu_{0,\eta}, \Sigma_{0,\eta})\,
   \phi(\gamma^{[g]} \mid \mu, \Sigma) \,
         p_{\mathrm{AUG}}(v^{[g]} \,|\,  y^{[g]}, \gamma^{[g]}, \eta ) \ .
\]

\emph{\textbf{Global parameters}}\\

Given realisations of the varying group-level parameters
$\gamma^{[1]}\ldots,\gamma^{[G]}$, global parameters
$\mu$ and $\Sigma$ can be updated using
Gibbs-sampling steps from the full conditional posteriors,
as explained in \citet{BDA3}, Section~3.6, and
\citet{HaganForster2004}, Chapter~14.
The conditional distribution of $\mu$ given  $\gamma^{[1]}\ldots,\gamma^{[G]}$, $\Sigma$
is given by
\[
\mathbf{\mu} \mid \mathbf{\Sigma},\gamma^{[1]}, \ldots, \gamma^{[G]}
  \sim \mathcal N_p
  \left( \frac{G}{\kappa_0 + G}\, \bar{\mathbf{\gamma}} +
 \frac{\kappa_0  }{\kappa_0 + G}\, \mathbf{\mu}_0,\ \frac{1}{\kappa_0+G}
      \, \mathbf{\Sigma} \right)
\]
with $\bar \gamma = (1/G)\sum_g \gamma^{[g]}$,
in which we recognize the posterior mean as a weighted sum of the
group-level parameters and the prior mean.

For the posterior variance-covariance matrix of $\gamma^{[g]}$ we have
\[
\mathbf{\Sigma} \mid \gamma^{[1]}, \ldots, \gamma^{[G]} \sim
         \mathrm{InvWishart}_{p_\gamma}(\mathbf{\Lambda}_1,\nu_0 + G),
\]
where
\begin{align*}
\mathbf{\Lambda}_1 & \,=\,
                  \mathbf{\Lambda}_0 + \mathbf{Q} +
                  \frac{\kappa_0 G}{\kappa_0 + G}
                  (\bar{\mathbf{\gamma}} - \mathbf{\mu}_0 )
                  (\bar{\mathbf{\gamma}} - \mathbf{\mu}_0)^{\prime} \ ,  \\
\mathbf{Q} & \,=\, \sum_{g=1}^G ( \gamma^{[g]} - \bar{\mathbf{\gamma}})
          ( \gamma^{[g]}- \bar{\mathbf{\gamma}})^{\prime} \  .
\end{align*}

The influence of the prior is mainly carried by $\mathbf{\Lambda}_0$ and the
last term of $\mathbf{\Lambda}_1$, which involves $\kappa_0$ and
$\mathbf{\mu}_0$.
Since the central tendency of the inverse Wishart($\Lambda, \nu$)
distribution is about $\nu^{-1}\Lambda$, this shows that
the posterior distribution of $\Sigma$ for large values of $G$
will be close to the variance-covariance matrix of the $\gamma^{[g]}$.

\emph{\textbf{Combining the updates}}\\

Sequentially the within-group ministeps $v$, the group-level parameters $\gamma$,
and the global parameters $\eta, \mu, \Sigma$ are updated.
To achieve good mixing, more updates are required for $v$ than for the other parameters.

\emph{\textbf{Initial values}}\\

Initial values are obtained in a procedure consisting of two stages.
First, parameters are estimated for the model where
all parameters in $\theta^{[g]}$ that are
coefficients in the linear predictor are assumed to be
constant across groups, but
the basic rate parameters are allowed to be group-dependent,
i.e., a multi-group model.
This estimation uses the Robbins-Monro algorithm proposed
for obtaining method-of-moments estimates in
\citet{Snijders01}, in a brief version because great precision
is not necessary here.
This yields an estimated value $\hat\theta^{(0)}$,
with estimated covariance matrix $C^{[0]}$. The components
of this vector and matrix corresponding to $\eta$
are denoted $\hat\eta^{(0)}$ and $C^{[0]}_\eta$.

Second, for each of the groups $g$ separately, starting from the
provisional estimate $\hat\theta^{(0)}$,
and keeping the components $\hat\eta^{(0)}$ constant,
a small number of Robbins-Monro steps again following \citet{Snijders01}
are taken to improve the estimate of $\theta^{[g]}$.
The result is used as initial value for $\hat\theta^{[g]}$.
The covariance matrix $C^{(1)}_g$ for the proposal distribution for $\theta^{[g]}$
is a weighted combination of the covariance matrix for this estimate and $C^{[0]}$.

\section{Data and model definition}

Data were collected in the first year of secondary school in
14 schools in the Netherlands in 2003-2004, with students
being on average slightly older than 12 years at the first wave.
There were four waves,
with three months in between. Allowing for the social processes to
be unstable at the very start of the school year, we used the last three waves.
These will be called waves 1-3 from now on,
which yields period~1 as the period from wave~1 to wave~2 and
period~2 as the period from wave 2 to~3.
Network $X$ was the friendship network, $Z$ the two-mode network of
delinquent behaviours with four second-mode nodes:
stealing, vandalism, graffiti, and fighting.

Covariates used were sex (female=1, male=2), language spoken at home,
and advice. The Dutch secondary school system is tiered and
`advice' here is defined as the recommended secondary school level
according to the advice given in the last grade of primary school.
It is ordered from low to high with range 1--9.

A basic measure for network stability in a period
is the Jaccard coefficient
\citep{BatBren95}, defined for network $X$ and period $m$ as
\[
\frac{\sum_{ij}  \min\{ x_{ij}(t_{m}), x_{ij}(t_{m+1})\}}
        {\sum_{ij} \max\{x_{ij}(t_{m}), x_{ij}(t_{m+1})\}}
\]
and for $Z$ similarly.

Delinquency was dichotomized to construct the two-mode network.
The coding was $z_{ih}=1$ if individual $i$ answered having
done the behaviour at least once in the past three months,
but for fighting the threshold was `at least twice' because apparently
this was rather common.

Of the original set of 126 classrooms, the criteria for including the classrooms
were having less than 20\% missing data in the first two waves for both
networks, but less than 10\% in the first wave for the delinquency network;
having at least 10 persons with non-missing advice;
and having Jaccard coefficients higher than 0.2 for both networks and both periods.
Furthermore, one group was excluded because it was considered an outlier
with a density for the delinquency network of more than 0.50 for all waves.
This leaves 81 groups.

\subsection{Model specification}

The mutual dependence between friendship and delinquent behaviour
was represented by the effects discussed in Section~\ref{S_interp}.
Here we discuss the effects operating only on the friendship
and those operating only on the delinquency network.
For the mathematical definition of the effects we refer to Appendix A.

The structural part of the model for friendship dynamics was defined
in accordance with what is usual for friendship networks.
The outdegree is a necessary effect, representing the balance between creation
and termination of ties.
Reciprocity and transitive triplets effects were included together
with their interaction following \citet{Block2015}.
As degree effects were included outdegree-activity, indegree-popularity,
and reciprocal degree-activity; for the latter a negative parameter
is expected, reflecting that actors
with more reciprocated friendship ties will tend to create
fewer new ties.
For the covariates we included homophily effects with respect to sex,
language, and advice, expecting positive parameters.
Furthermore, the logarithm of group size was included to account for
group size differences, where a parameter in the neighbourhood of $-0.5$
was expected.

For the delinquency network effects included were the outdegree effect,
outdegree-activity and indegree-popularity reflecting, respectively,
differences between students and between delinquent activities,
and effects of sex, advice, and classroom mean advice.

For this multilevel network model with Bayesian estimation,
it was mentioned above that it is possible to specify fairly many parameters
as randomly varying between groups, but not too many.
In any case, the rate parameters must vary randomly between groups.
A moderate number of random effects was chosen.
Random effects were given to outdegree, reciprocity, indegree-popularity,
reciprocal degree-activity, transitivity, same language,
and similar advice for the friendship network;
and to outdegree and outdegree-activity for the delinquency network.

\subsection{Prior specification}

For the rate parameters a data-dependent normal prior was used,
with means and covariance matrices given by the robust mean 
and $0.5$ times the covariance matrix 
of the rate parameter estimates in the multi-group estimation.

For the parameters of the evaluation function, the determination of the
prior distribution was based on existing
experience with modeling friendship networks,
together with the desire not to influence the results too strongly,
while still obtaining convergence of the MCMC process.
It should be noted that non-zero prior means chosen for structural
parameters below have little influence on the results.
The evidence for reciprocity, for example, is typically strong enough
to overwhelm the prior (see Appendix B for a brief illustration) and
the performance of the MCMC is typically not contingent on a strong prior.
Naturally, it would be unwise to choose a strong informative prior for
any parameter that is the main target of inference.

The effects used in this example all are scaled in such a way that their
parameters have sizes usually between $-1$ and $+1$, except for
the outdegree parameter which is negative, reflecting the sparsity
of the networks, and reciprocity, which often has a parameter between +1 and +3.
This implies a prior uncertainty of the global means $\mu$
with a standard deviation of approximately 1; for the outdegree
parameters the prior uncertainty is larger.
Furthermore, the groups will tend to be similar to each other,
which we express by the prior expectation that the between-group
standard deviations are 10 times smaller than the prior standard
deviations for the elements of $\mu$.
This is reflected by the value $\kappa_0 = 0.01$.

These considerations led to prior means of $-2$ for the
outdegree parameters, $+1$ for reciprocity,
$+0.2$ for transitive triplets, and 0 for all other
coefficients of random effects.
For the 13-dimensional prior Wishart distribution,
$\nu_0^{-1}\Lambda_0$ was chosen as
a diagonal matrix with diagonal values 0.01
except for the two outdegree parameters, which had value 0.1;
and number of degrees of freedom $\nu_0 = 15$.

For the fixed effects, improper constant prior distributions
were used for all except the effects of the
group-level variables which are log group size and group mean of advice;
for these parameters the prior distribution was normal 
with mean~0 and variance~0.04.

\section{Results}

\subsection{Descriptive statistics}

Table~\ref{T_delinquentActs} gives the overall means and Jaccard similarity
coefficients of the four delinquent acts for the pooled data.
They are positively associated.

\begin{table}
\caption{\label{T_delinquentActs}Overall means and similarity coefficients of delinquent acts}
\centering
\begin{tabular}{l|c| ccccc|}
\cline{2-7}
&\emph{mean}  & & \multicolumn{4}{c|}{\emph{Jaccard similarity}} \\
   &  & \hspace{2em} & stealing   & vandalism & graffiti & fighting \\
stealing   & 0.13 &  & --   & 0.29 & 0.22 & 0.27 \\
vandalism  & 0.20 &  & 0.29 &  --  & 0.27 & 0.33 \\
graffiti   & 0.17 &  & 0.22 & 0.27 &  --  & 0.23  \\
fighting   & 0.21 &  & 0.27 & 0.33 & 0.23 &  --  \\
\cline{2-7}
\end{tabular}
\end{table}

A measure for delinquency is the outdegree in the two-mode network, i.e.,
the number of delinquent behaviours reported by a student.
For this variable and for the covariates the means,
within-classroom and between-classroom standard deviations
($\hat\sigma$ and $\hat\tau$), and the
intraclass correlation coefficients (icc)
\citep[calculated according to][Chapter 3]{SnijdersBosker12}
are reported in Table~\ref{T_descr_vars}.
From the intraclass correlation coefficients we see that
the classrooms are quite homogeneous with respect to advice,
not assortative with respect to sex,
while for the level of delinquency and whether the Dutch language
is spoken at home assortativity is positive but low.

\begin{table}
\caption{\label{T_descr_vars}Descriptives for actor variables}
\centering
\begin{tabular}{l  r@{.}l  r@{.}l  r@{.}l  r@{.}l}
\hline
    & \multicolumn{2}{c}{mean} &  \multicolumn{2}{c}{$\hat\sigma$}  &
 \multicolumn{2}{c}{$ \hat\tau$}   &  \multicolumn{2}{c}{icc} \\
sex M              & 0&53  & 0&50  & 0&03    & 0&00    \\
advice             & 6&71  & 0&88  & 1&44    & 0&73    \\
language Dutch     & 0&91  & 0&27  & 0&07    & 0&06    \\
delinquency wave 1 & 0&76  & 1&03  & 0&21    & 0&04    \\
delinquency wave 2 & 0&90  & 1&10  & 0&25    & 0&05    \\
delinquency wave 3 & 0&90  & 1&14  & 0&27    & 0&05    \\
\hline
\end{tabular}
\end{table}

Some descriptive statistics for the set of 81 friendship networks
are presented in Table~\ref{T_descr_fr}.
These include reciprocity, defined as the proportion of ties $i \rightarrow j$
that is reciprocated by $j \rightarrow i$, and transitivity,
defined as the proportion of two-paths $i \rightarrow j \rightarrow h$
that is closed by $i \rightarrow h$.
Average degrees are about 4, average reciprocity is about 0.60,
and average transitivity is about 0.56.
These are quite usual figures for friendship networks.
The Jaccard measure for network stability ranges for
friendship from 0.28 to 0.75, with a mean of 0.51.
This indicates that a good proportion of ties remains in place
from one wave to the next.

\begin{table}
\caption{\label{T_descr_fr}Descriptives for friendship networks: means and standard deviations
  across the 81 groups.}
\centering
\begin{tabular}{l  r@{.}l  r@{.}l  r@{.}l  r@{.}l  r@{.}l  r@{.}l}
\hline
    & \multicolumn{4}{c}{wave 1} &  \multicolumn{4}{c}{wave 2}  &
     \multicolumn{4}{c}{wave 3} \\
    & \multicolumn{2}{c}{mean} &  \multicolumn{2}{c}{(s.d.)}
    & \multicolumn{2}{c}{mean} &  \multicolumn{2}{c}{(s.d.)}  &
     \multicolumn{2}{c}{mean} &  \multicolumn{2}{c}{(s.d.)}  \\
mean outdegree   & 4&01  & (0&66)\phantom{x} & 4&17    & (0&60)\phantom{x}  & 4&03    & (0&68) \\
 s.d.\ outdegree  & 2&62  & (0&88)  & 2&72    & (0&73)  & 2&55    & (0&73) \\
 s.d. indegree    & 1&96  & (0&54)  & 1&97    & (0&57)  & 1&99    & (0&50) \\
 reciprocity      & 0&59  & (0&08)  & 0&60    & (0&09)  & 0&60    & (0&09) \\
transitivity     & 0&55  & (0&09)  & 0&56    & (0&09)  & 0&56    & (0&09) \\
Jaccard with next wave & 0&50 & (0&09) & 0&52 & (0&08)  &  \multicolumn{4}{c}{ } \\
proportion missings & 0&03  & (0&03)  & 0&07    & (0&06)  & 0&06    & (0&04) \\
\hline
\end{tabular}
\end{table}

Some descriptive statistics for the two-mode delinquency networks
are given in Table~\ref{T_descr_del}.
The students report on average slightly less than one out of the
four delinquent acts.
The Jaccard measure for stability ranges from 0.21 to 0.70, with a mean of 0.41.
Here also there is some change from one wave to the next, but not too much.

\begin{table}
\caption{\label{T_descr_del}Descriptives for delinquency networks: means and standard deviations
  across the 81 groups.}
\centering
\begin{tabular}{l  r@{.}l  r@{.}l  r@{.}l  r@{.}l  r@{.}l  r@{.}l}
\hline
    & \multicolumn{4}{c}{wave 1} &  \multicolumn{4}{c}{wave 2}
    & \multicolumn{4}{c}{wave 3} \\
    & \multicolumn{2}{c}{mean} &  \multicolumn{2}{c}{(s.d.)}
    & \multicolumn{2}{c}{mean} &  \multicolumn{2}{c}{(s.d.)}
    & \multicolumn{2}{c}{mean} &  \multicolumn{2}{c}{(s.d.)}  \\
mean outdegree   & 0&76  & (0&29) \phantom{x} & 0&91    & (0&34)\phantom{x}  & 0&92    & (0&35) \\
s.d.\ outdegree  & 1&01  & (0&23)  & 1&09    & (0&20)  & 1&12    & (0&24) \\
s.d. indegree    & 2&02  & (0&83)  & 2&14    & (1&01)  & 2&01    & (1&00) \\
Jaccard with next wave & 0&39 & (0&09) & 0&43 & (0&10) &  \multicolumn{4}{c}{ } \\
proportion missings & 0&01  & (0&02)  & 0&06    & (0&05)  & 0&06    & (0&05) \\
\hline
\end{tabular}
\end{table}

\subsection{Modelling results}

For the MCMC procedure, three parallel chains were used, each of 70,000 steps;
each step consisted of 200-800 updates of $v$ in the 81 groups (with a total of 66,200),
five updates of $\eta$, and one update of each of $\gamma$, $\mu$, and $\Sigma$.
Of the 70,000, the first 10,000 were the warming phase.
The homogeneity of the three chains was good according to the $\hat R$ measure of
\citet{BDA3}, which was less than 1.05 for all global parameters.

Posterior means, standard deviations, and credibility intervals of the parameters
are given in Table~\ref{T_results}.
Appendix~C illustrates the posterior distributions of some parameters.
The estimated model for friendship dynamics is usual and has the
usual interpretation \citep[e.g.,][]{FujimotoEA2018,SienaManual21}.
We focus the interpretation on the mutual dependency of delinquency
and friendship, using the mnemonic indicators given above in the
list of cross-network effects.

\noindent
\emph{Dependent variable: friendship}\\
The delinquency degree, i.e., the number of delinquent acts
practised, is a measure for delinquent behaviour.
Effects of delinquent behaviour on friendship dynamics are minor.
The table shows that delinquency has virtually no effect on the
popularity as a friend (`id') and a negative effect on the activity
in nominating friends (`od'): those who report more delinquent behaviour
tend to nominate slightly fewer friends.
Practising the same delinquent acts (`odd') has no appreciable effect
on friendship formation.

\begin{table}
\caption{\label{T_results}Posterior summaries for delinquency networks}
\centering
\begin{tabular}{l | r@{.}l r@{.}l r@{.}l r@{.}l r@{.}l | }
\hline
\rule{0pt}{2ex}\relax
Effect &\multicolumn{2}{c|}{par.}&\multicolumn{2}{c }{ (psd) } &\multicolumn{4}{c }{ CI } &
            \multicolumn{2}{c | }{betw.\ sd} \\[0.5ex]
\hline
\multicolumn{5}{l}{\emph{friendship}}&\multicolumn{2}{c}{0.025}&\multicolumn{2}{c}{0.975}&\multicolumn{2}{c}{}\\
\hline
\rule{0pt}{2ex}\relax
outdegree (density)          & --2 & 325 & (0 & 068) & --2 & 46 & --2 & 19 & 0 & 386\\
reciprocity                  &   2 & 044 & (0 & 061) & 1 & 93 &  2 & 16 & 0 & 332\\
transitive triplets          &   0 & 457 & (0 & 015) & 0 & 43 &  0 & 49 & 0 & 100\\
transitive recipr. triplets  & --0 & 149 & (0 & 016) & --0 & 18 & --0 & 12 &  \multicolumn{2}{c|}{ } \\
indegree - popularity        & --0 & 074 & (0 & 012) & --0 & 10 & --0 & 05 & 0 & 092\\
outdegree - activity         &   0 & 038 & (0 & 004) & 0 & 03 &  0 & 05 & \multicolumn{2}{c|}{ } \\
reciprocal degree - activity & --0 & 186 & (0 & 014) & --0 & 21 & --0 & 16 & 0 & 085\\
same sex                     &   0 & 660 & (0 & 024) & 0 & 61 &  0 & 71 & \multicolumn{2}{c|}{ }\\
log class size               & --0 & 127 & (0 & 166) & --0 & 46 &  0 & 20 & \multicolumn{2}{c|}{ }\\
advice similarity            &   0 & 105 & (0 & 084) & --0 & 06  & 0 & 27 & 0 & 250\\
same language                &   0 & 172 & (0 & 033) & 0 & 11 & 0 & 24 & 0 & 172\\
delinq.\ degree popularity `\id'&--0& 001 & (0 & 013) & --0 & 03 & 0 & 02 & \multicolumn{2}{c|}{ }\\
delinq.\ degree activity `\od'& --0 & 043 & (0 & 013) & --0 & 07 & --0 & 018 & \multicolumn{2}{c|}{ }\\
same delinquent acts `\odd'   &   0 & 040 & (0 & 030) & --0 & 02 &  0 & 10 & \multicolumn{2}{c|}{ }\\
\hline
\multicolumn{9}{l}{\emph{delinquency}}\\
\hline
\rule{0pt}{2ex}\relax
outdegree (density)              & --2 & 250 & (0 & 087) & --2 & 41 & --2 & 08 & 0 & 333\\
indegree - popularity            &   0 & 012 & (0 & 011) & --0 & 01  & 0 & 03 & \multicolumn{2}{c|}{ }\\
outdegree - activity             &   0 & 406 & (0 & 018) & 0 & 37  & 0 & 44 & 0 & 098\\
sex (M)                          &   0 & 207 & (0 & 042) & 0 & 13 &  0 & 29 & \multicolumn{2}{c|}{ }\\
advice                           &   0 & 018 & (0 & 022) & --0 & 02 &  0 & 06 & \multicolumn{2}{c|}{ }\\
classroom mean advice            & --0 & 042 & (0 & 031) & --0 & 10 &  0 & 02 & \multicolumn{2}{c|}{ }\\
friendship indegree activity `\id'& --0 & 004 & (0 & 012) & --0 & 03 &  0 & 02 & \multicolumn{2}{c|}{ }\\
friendship outdegree activity `\od'&--0 & 073 & (0 & 015) & --0 & 10 & --0 & 04 & \multicolumn{2}{c|}{ }\\
same delinq.\ acts as friends `\odd'& 0 & 267 & (0 & 037) &0 & 19  & 0 & 34 & \multicolumn{2}{c|}{ }\\
av.\ number of delinq.\ acts of friends `\odp' & --0 & 125 & (0 & 051) &   --0 & 22 & --0 & 02 & \multicolumn{2}{c|}{ }\\
\hline
\multicolumn{9}{l}
   {\footnotesize{par = posterior mean $\hat\mu, \hat\eta$;
   psd = posterior standard deviation of $\mu, \eta$;}}\\
\multicolumn{9}{l}
   {\footnotesize{betw.\ sd = posterior
        between-groups standard deviation $\hat\sigma$.}}\\
\end{tabular}
\end{table}

\noindent
\emph{Dependent variable: delinquent acts}\\
We start with discussing the five effects not related to friendship.
The parameter for the outdegree effect
($\hat\mu_k = -2.250$) indicates a reluctance to practising
delinquency, stronger for girls than for boys ($\hat\eta_k = 0.207$).
There is hardly a differentiation between the four
delinquent acts (indegree popularity, $\hat\eta_k = 0.012$)
but quite a strong differentiation between students
(outdegree activity, $\hat\mu_k = 0.406$), expressing that those
currently practising more delinquency have a stronger tendency
to add new delinquent acts. The evidence is inconclusive for effects of school advice,
and of its classroom mean (the latter is negative with $0.91$ posterior probability).

Social influence is represented by the four
mixed effects of friendship and delinquency on
the dynamics of delinquent behaviour.
The effects of indegrees (`id') and outdegrees (`od')
show a similar pattern to what was found for friendship dynamics:
there is a negative effect of the outdegree for friendship
on the number of delinquent acts reported.
There is a rather strong tendency to practise the same delinquent
acts as one's friends (`odd', $\hat\eta_k = 0.267$)
but a negative effect of the average
delinquency of friends (`od\_av', $\hat\eta_k = -0.125$).

The combination means that, for a given delinquent act $h$,
if more of $i$'s friend practise it then $i$ will have a higher
probability of also starting to practise it and a lower probability of stopping with it;
by contrast, if $i$'s friends are more delinquent on average but none of them
practises act $h$ then $i$ will have a lower probability of starting
to practise $h$, and a higher probability of stopping.
However, the former, positive influence effect is stronger
than the latter, negative influence effect, because its parameter
is higher in absolute value and the range of the explanatory variable
corresponding to `odd' --- which is the
number of friends practising act $h$ --- is equal to 9, which is larger than
the within-group range of the explanatory variable corresponding to `od\_av', equal to 4.

Concluding, there is a weak social selection effect, where those
who are more delinquent tend to nominate fewer friends,
and a rather strong social influence effect in the sense of
practising the same delinquent behaviours as one's friends, but
a weak effect of avoiding the delinquent behaviours not
practised by one's friends if the friends are more delinquent otherwise.
This contrasts with the results of \citet{KnechtEA2010},
who used the same data set but found no evidence of social influence.
That publication used a simpler two-stage multilevel network method which
allowed the inclusion of only 21 classrooms of this data set.
Another difference is that the earlier publication did not distinguish
the four separate delinquent acts in a two-mode network, but used
an aggregate measure of delinquency. This leads to incomparability between
the data used in the analysis, because the two-mode network representation
required dichotomization of the four delinquency variables,
while they were added, without dichotomization, in \citet{KnechtEA2010}.

\section{Conclusions}
Network analysis has typically been concerned with describing and modelling
network processes for individual networks only. We have proposed a modelling
framework for generalising network inference beyond the specifics of individual
groups to a population of networks.
The model is a hierarchical extension of the Stochastic Actor-oriented Model
\citep{Snijders2017} for longitudinal network panel data, using random coefficients
to represent differences between groups.
This allows taking into consideration group-level effects, e.g.,
interventions or compositional characteristics,
and their cross-level interactions with within-group effects.
A further possibility is to investigate the network dynamics in many small groups,
e.g., of size 5 to 10, for which an analysis per group does not give
meaningful results; an example is \citet{Dolgova2019}.

The methods are implemented in the R package RSiena \citep{SienaManual21}.
They have been available in beta versions since a few years, which already led
to applications, e.g., \citet{Boda2018}.
The MCMC algorithm proposed in this paper is a straightforward procedure,
and future work will be devoted to making it more efficient.

\section*{Acknowledgements}
This work was supported in part by award R01HD052887 from the US Eunice Kennedy Shriver National Institute of Child Health and Human Development (John M. Light, Principal Investigator). We are grateful to Ruth Ripley and her programming and support in the foundational stages of this project at Nuffield College and the Department of Statistics at Oxford.

\medskip

\bibliography{MultilevelSAOM_2022.bib}

\newpage
\appendix

\centerline{\textbf{\Large Appendix}}

\section{Statistics}
A comprehensive list and definition of all effects currently employed in SAOMs
is provided in \citet[][Chapter 12]{SienaManual21}.
The effects used in this paper are defined as follows.
\begin{enumerate}
    \item Outdegree (density)\\
    $s_{\mathrm{d},i}^X (x)= \sum_j x_{ij}$
     \item Reciprocity \\
    $s_{\mathrm{rec},i}^X (x)= \sum_j x_{ij} \, x_{ji}$
     \item transitive triplets\\
    $s_{\mathrm{tt},i}^X (x)= \sum_{j,h} x_{ij} \, x_{ih} \, x_{hj}$
     \item transitive reciprocated triplets effect\\
    $s_{\mathrm{trt},i}^X (x)= \sum_{j,h} x_{ij \,} x_{ji} \, x_{ih} \, x_{hj}$
    \item indegree-popularity \\
    $s_{\mathrm{idp},i}^X (x)= \sum_{j,h} x_{ij} \, x_{hi} $
    \item outdegree-activity \\
    $s_{\mathrm{oda},i}^X (x)= \sum_{j,h} x_{ij} \, x_{ih} $
    \item reciprocal degree activity \\
    $s_{\mathrm{rda},i}^X (x)= \sum_{j,h} x_{ij} \, x_{ih} \, x_{hi} $
    \item covariate $B$ ego \\
    $s_{\mathrm{ego},i}^X (x)= \sum_{j} x_{ij} \, B_i $
    \item same covariate $B$ \\
    $s_{\mathrm{same},i}^X (x)= \sum_{j} x_{ij} \, I\{ B_i = B_j\} $
    \item covariate $B$ similarity \\
    $s_{\mathrm{sim},i}^X (x)= \sum_{j} x_{ij} \,
           \big( c - | B_i - B_j | / \mathrm{range}(B)  \big) $, \\
    where $c$ is a centering constant
\end{enumerate}
The effects for network $Z$ are similar. The cross-network effects were defined
in Section~\ref{S_interp}. The effect of covariates in Table~\ref{T_results}
are ego effects, unless indicated as `same' or `similarity'.

\section{Prior sensitivity}

\subsection{Prior variance}
\begin{figure}[tp]
\begin{center}
\includegraphics[scale=0.5]{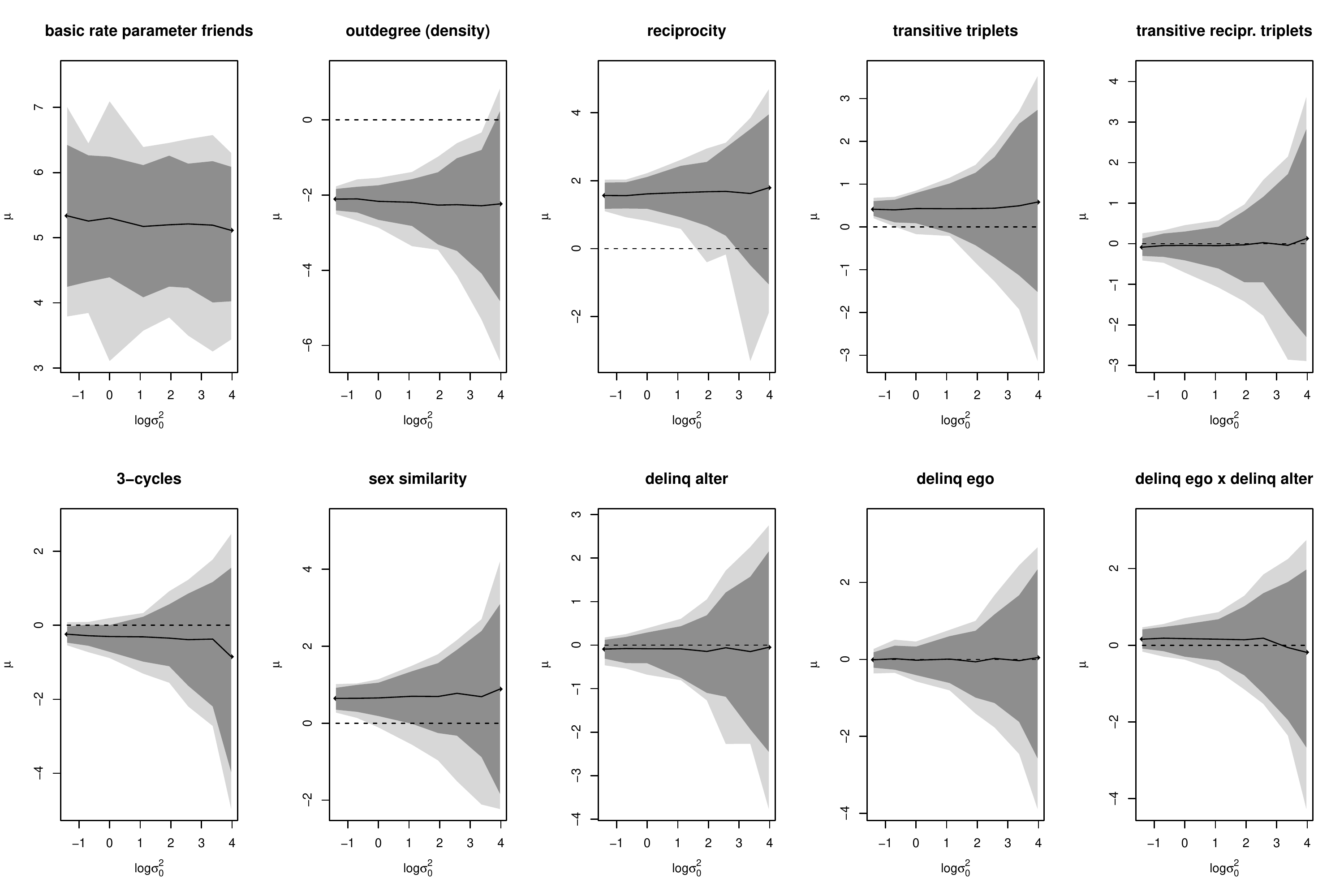}
\end{center}
\caption{Credibility intervals (95\%, dark; 99\% light) for $\mu$ with default prior and $\mathbf{\Lambda}_0=\sigma^2_0 I$ for different values of $\log(\sigma^2_0)$}
\label{fig:shrinkageprior}
\end{figure}

As outlined in Section 6, the influence of the prior is mainly from $\mathbf{\Lambda}_0$
and will affect the inference both for $\mu$ and the group-wise parameters $\gamma^{[g]}$,
through $\Sigma$. As an illustrative example of the prior scale,
we consider here the subset of 21 classrooms used in \citet{KnechtEA2010} 
for a simplified model for the network only, $\mathcal{Y}^{[g]}=\mathcal{X}^{[g]}$
($g=1,\ldots,21$) and $M=2$ periods. In the structural part,
we have omitted indegree popularity, outdegree activity, and reciprocity activity
but added 3-cycles. To further simplify the model, delinquency is treated
as a nodal covariate. All parameters are variable and $p_2=0$. Figure~\ref{fig:shrinkageprior} provides the credibility intervals for $\mu_k$
using the Normal - Inverse Wishart prior with $\mu_0=0$, $\nu_0=12$,
$\kappa_0=1$, $\mathbf{\Lambda}_0=\sigma^2_0 I$,
for different values of $\sigma^2_0$
(a very small number of draws, 300, have been used here).

For small values of  $\sigma^2_0$, the credibility intervals are noticeably
tighter than for increasingly large values $\sigma^2_0$, when the prior
variance overwhelms the data.
The central tendencies (posterior means) are remarkably constant as a function of  $\sigma^2_0$, and are hardly pulled towards the prior mean of zero,
even for values of $\sigma^2_0$ as small as $0.25$ (the smallest value in the plots).

\begin{figure}[tp]
\begin{center}
\includegraphics[scale=0.5]{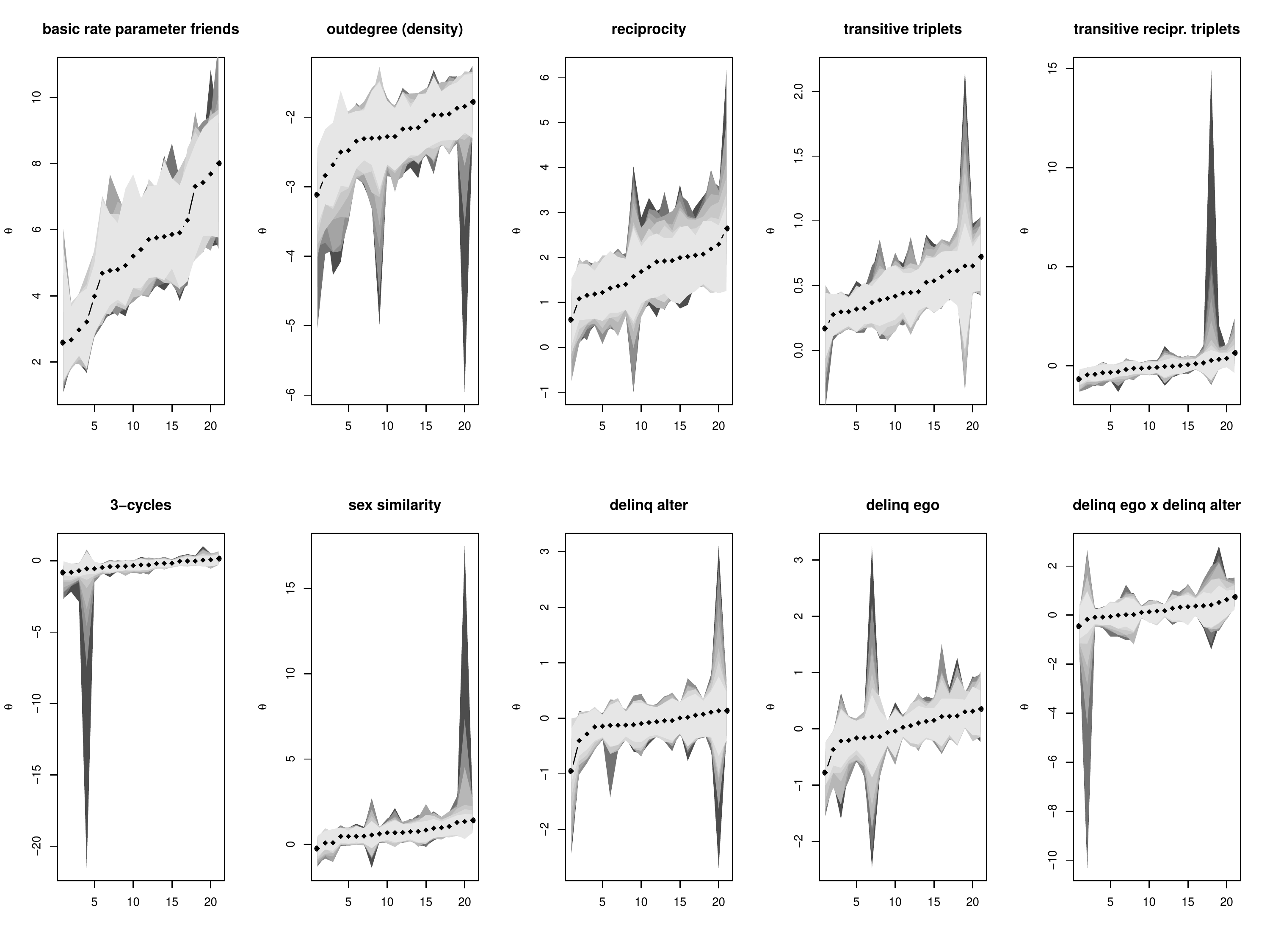}
\end{center}
\caption{Equal 95\% tail prediction intervals for $\gamma^{[g]}$ for different values of $\sigma^2_0$, from $\sigma^2_0=0.25$ (light grey) to $\sigma^2_0=113$ (dark grey). Groups ordered according to posterior predictive mean}
\label{fig:shrinkageprediction}
\end{figure}

The influence on the group parameters $\gamma^{[g]}_k$  of the same set of priors
is illustrated in Figure~\ref{fig:shrinkageprediction}.
Note the difference in vertical scale.
The inference on these parameters is remarkably robust to the prior variance.
Only for extreme values of $\sigma^2_0$ and for a few classrooms 
do we see a big change in group-level parameters. 
The very wide intervals are due to two specific classrooms.
More specifically, in one classroom (number 20) 
`transitive reciprocated triples' and `3-cycles' were collinear, 
which manifests itself in extremely large intervals 
for these parameters when large $\sigma^2_0$
prevents this classroom from borrowing information from the other classrooms.
Another classroom (number 11) had a similar issue with structural parameters
and in addition a `sex similarity' effect that is difficult to estimate
because of the very skewed sex distribution in this classroom.
The issues with these two classrooms also manifests themselves in
increasingly poor mixing for $\gamma^{[g]}$ for
large values of $\sigma^2_0$ (results available upon request from the authors).

\subsection{Reference prior}

We may consider the influence of the prior for $\mu$ and $\Sigma$ on the predictive distributions for $\gamma^{[g]}$ by comparing these to posteriors from group-level parameters estimated independently.
The \textit{a priori} dependence of $\mu$ on $\Sigma$ can be decoupled by setting $\pi(\mu,\Sigma)=\pi(\mu)\pi(\Sigma)$. When $G>p_1+1$, we may chose an improper prior for $(\mu,\Sigma)$ for reference.

Jeffreys rule \citep[][]{jeffreys1998theory} is a principled choice for a reference prior. For the multivariate normal distribution, it is given by
\[
p(\mu,\Sigma) \propto |\Sigma|^{-(p_1+2)/2}{\text{ ,}}
\]
or (for the independence-Jeffreys prior)
\[
p(\mu,\Sigma) \propto |\Sigma|^{-(p_1+1)/2}{\text{ .}}
\] 
For the conjugate model this corresponds to $\kappa_0 \rightarrow 0$, $\nu_0 \rightarrow 0$, and letting the determinant of $\Lambda_0$ tend to $0$.
Jeffreys prior is still conjugate for $\mu$ and $\Sigma$, and as such does not alter the updating scheme outlined in Section~\ref{S_estimation}.

\begin{figure}[tp]
\begin{center}
\includegraphics[scale=0.5]{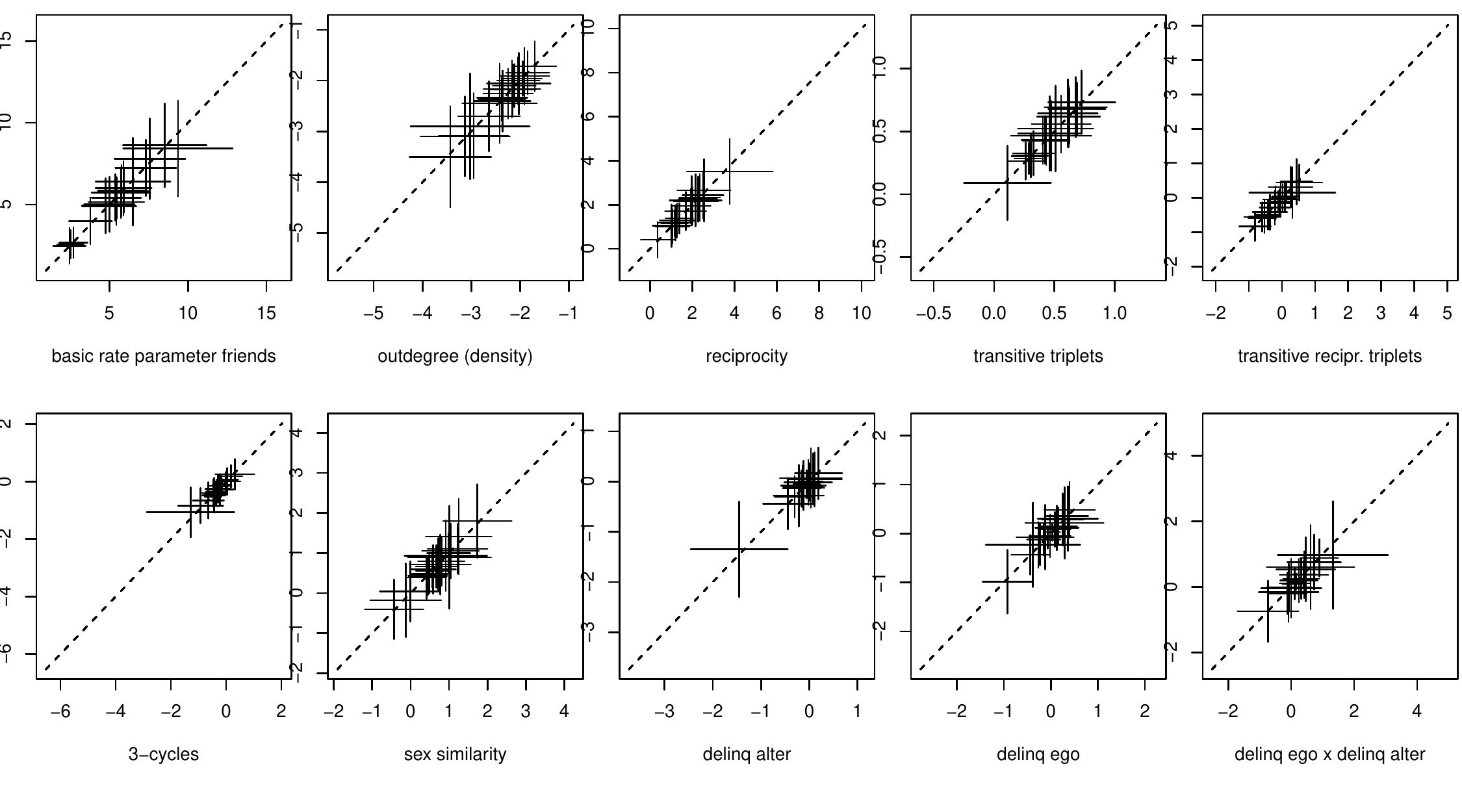}
\end{center}
\caption{Prediction-intervals for $\gamma^{[g]}$ fitted independently (horizontal axis) against predictions from Hierarchical SAOM using Jeffreys prior (vertical)
(excluding classrooms $g=11,20$)}
\label{fig:jeffpred}
\end{figure}

Figure~\ref{fig:jeffpred} illustrates the inference obtained from (horizontal)
fitting the model separately to each classroom, assuming a constant prior, and
(vertical) the predictive distributions obtained from the
hierarchical SAOM with Jeffreys prior.
The two previously mentioned classrooms 11 and 20 are omitted for reasons mentioned above.
Both analyses are based on the other 19 classrooms.\\

\begin{figure}[tp]
\begin{center}
\includegraphics[scale=0.5]{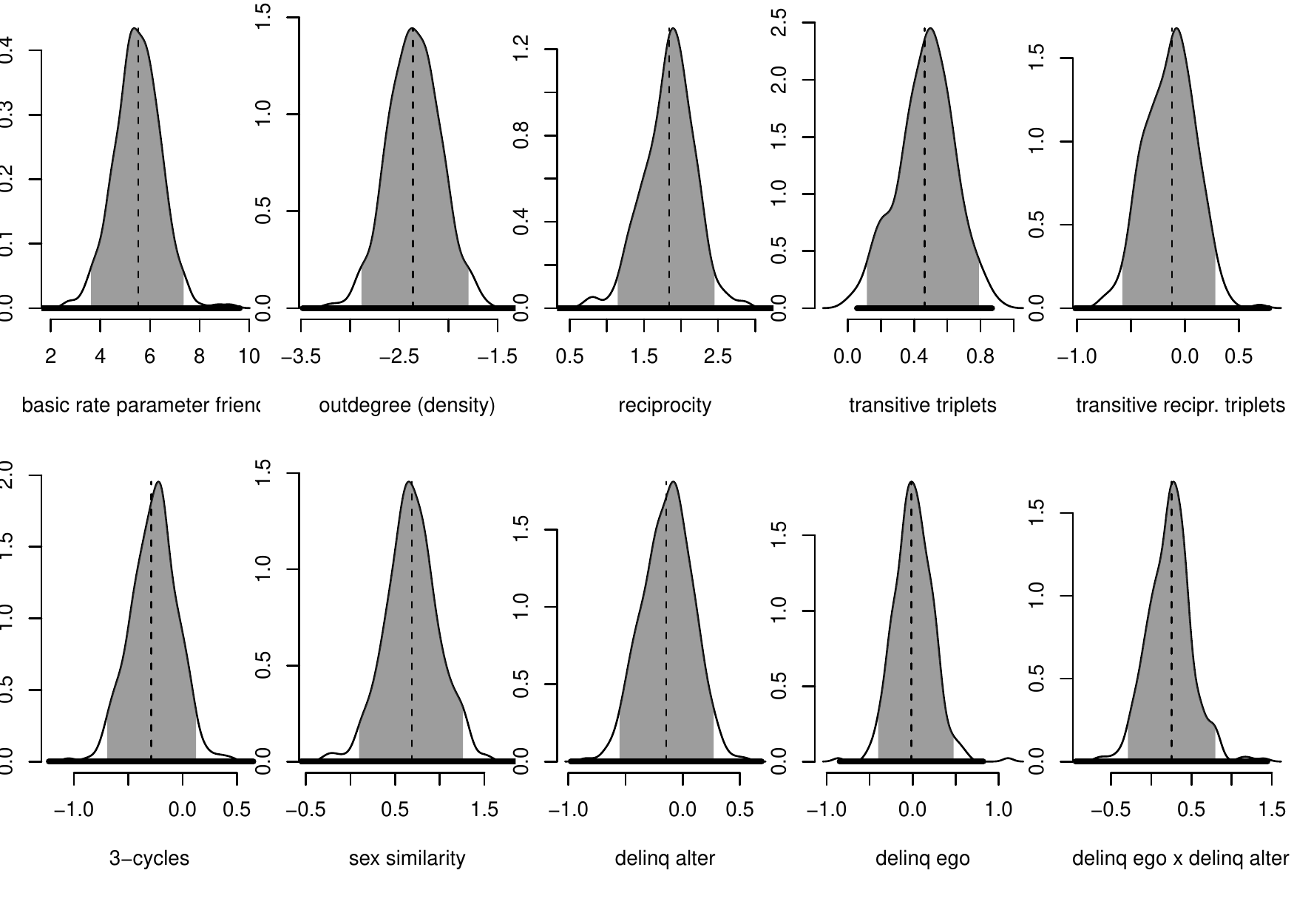}
\end{center}
\caption{Posteriors for $\mu$ fitted using Jeffreys prior with vertical line representing $\bar{\gamma}_k$ and the horizontal bar $\bar{\gamma}_k\pm 2 sd(\gamma^{[g]}_k)$ from independently fitting each group (excluding rogue $g=11,20$)}
\label{fig:jeffmu}
\end{figure}

Figure~\ref{fig:jeffpred} demonstrates a negligible influence of the prior 
on the distributions for $\gamma^{[g]}$.
Figure~\ref{fig:jeffmu} presents the posterior densities for $\mu_k$
and shows that these are also centered on the raw, un-weighted means
$\bar \gamma_k$ of $\gamma^{[g]}_k$ from the separate estimations.
This shows that imposing the multivariate normal model for the group-level parameters does not yield results that differ strongly 
from the individual group-level inference.
In order to make use of all classrooms we would however require more
additional classrooms to borrow strength across groups,
and impose a more informative prior for $\mu$ and $\Sigma$.


\section{Posteriors}
\begin{figure}[tp]
\begin{center}
\includegraphics[scale=0.8]{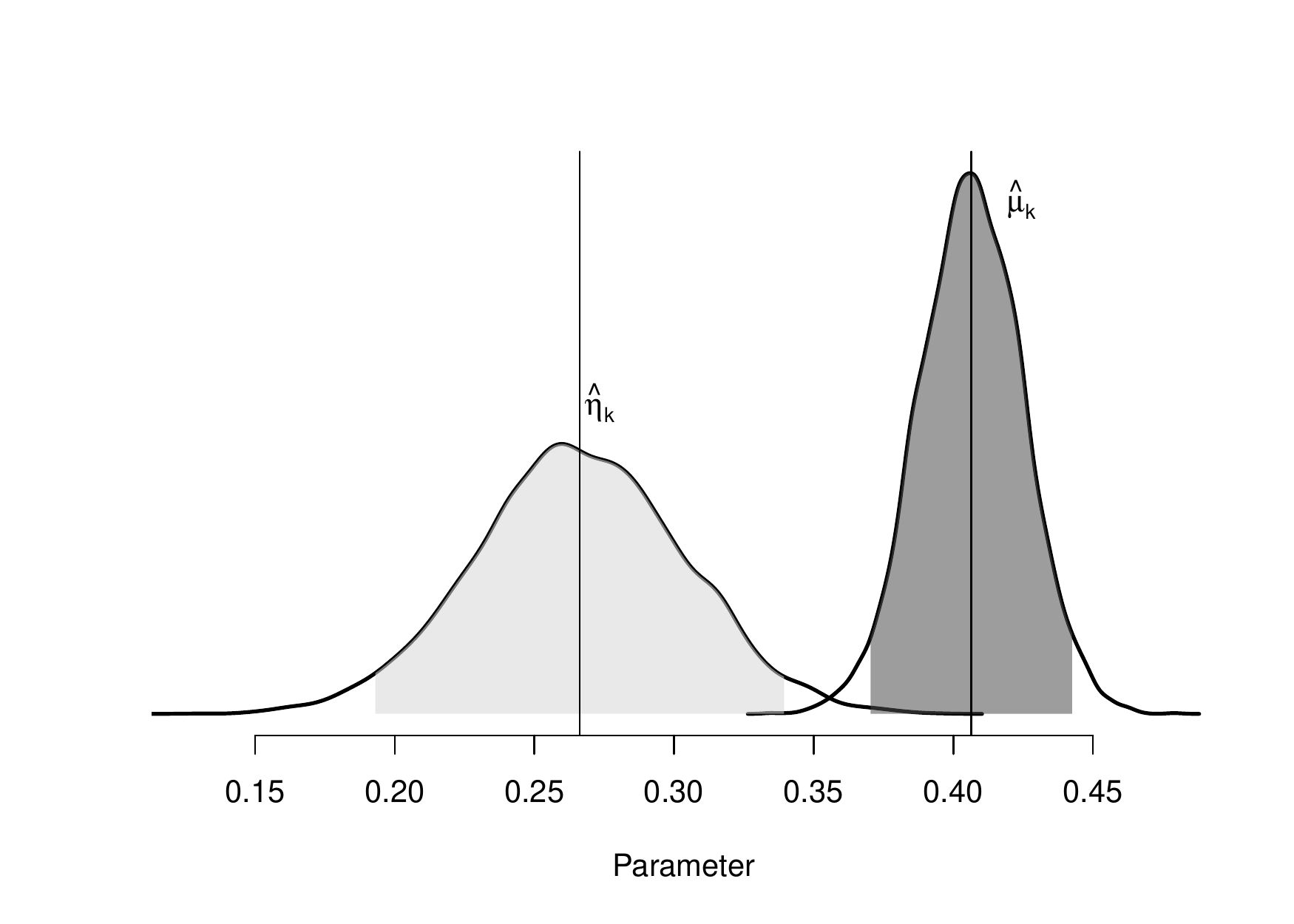}
\end{center}
\caption{Posterior densities for $\mu_k$ for delinquency outdegree - activity and $\eta_k$ for same delinquency acts as friends (`odd') with 95\% credibility intervals in dark and light grey, respectively.}
\label{fig:posteriormueta}
\end{figure}

\begin{figure}[tp]
\begin{center}
\includegraphics[scale=0.85]{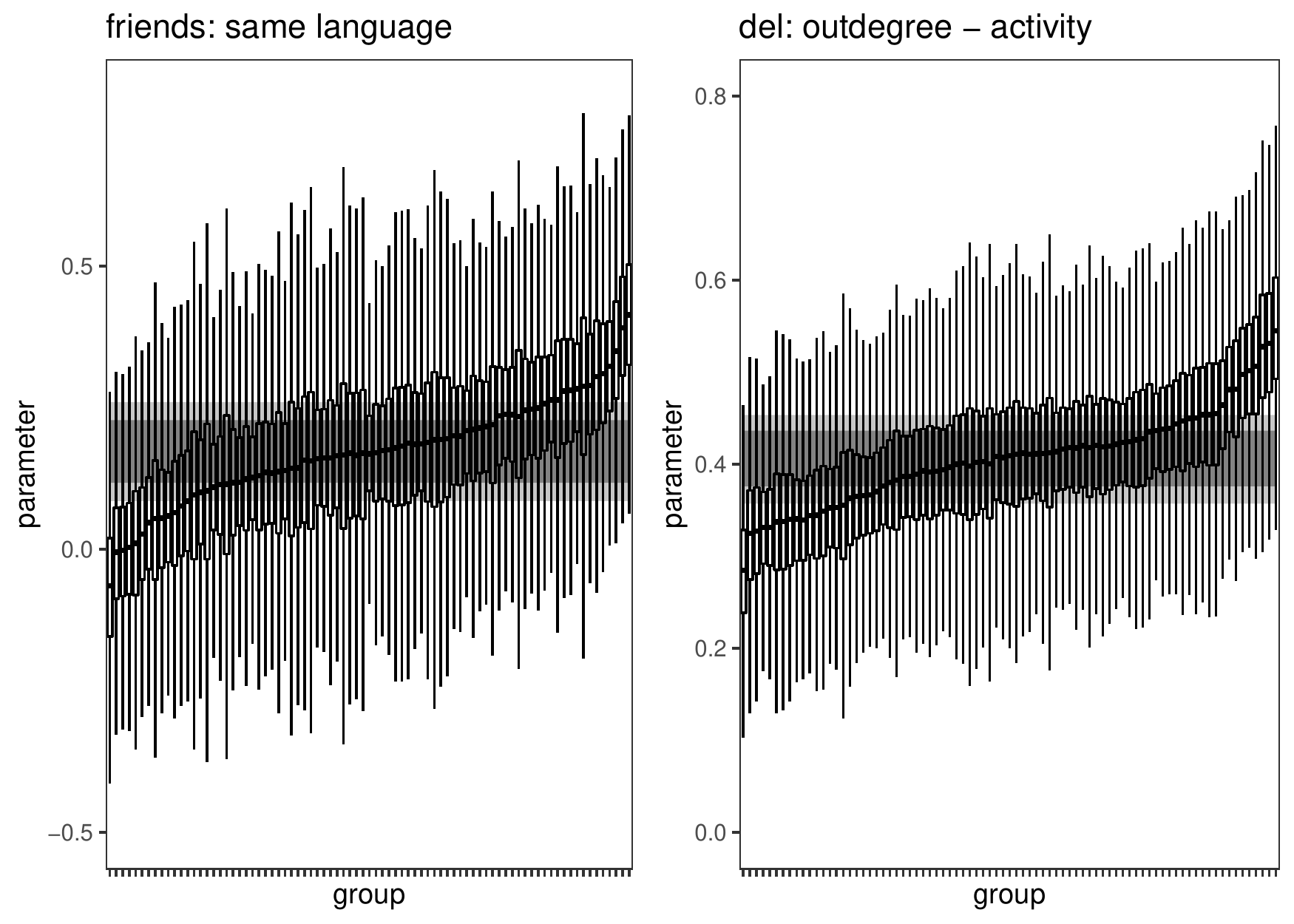}
\end{center}
\caption{Boxplots (without outliers) for posteriors of $\gamma_k^{[g]}$
 for same language and delinquency outdegree - activity,
 ordered by posterior means.
 The horizontal dark and light grey bands indicate the 90\% and 99\%
 credibility intervals for $\mu_k$.}
\label{fig:posteriormugamma}
\end{figure}

For the estimated model of Table~\ref{T_results},
Figure~\ref{fig:posteriormueta} presents the posterior distributions of the
population mean $\mu_k$ for the delinquency outdegree - activity effect and
of the constant parameter $\eta_k$ for the effect (`odd') 
of same delinquency acts as friends. 
For both parameters it is evident that they are positive with a 
high posterior probability. 
There is less posterior uncertainty about $\mu_k$ than $\eta_k$ but
the former is a population mean of varying group-wise parameters 
$\gamma_k^{[g]}$ whereas the latter is a constant parameter.
This variability across groups is illustrated in Figure~\ref{fig:posteriormugamma} (right panel), which shows 
boxplots (without outliers) of the posterior distributions of 
$\gamma_k^{[g]}$ for groups $g$ ordered according to the posterior means,
with horizontal credibility bands in grey for $\mu_k$.
The variability in group-level means is greater than the 
variability in $\mu_k$ but $\gamma^{[g]}$ is clearly positive 
with high posterior probability for all groups $g$. 
The length of the 95\%CI for delinquency outdegree - activity is~$0.072$ 
and the length of the 95\%CI for same language is~$0.132$ 
but both parameters are positive with high posterior probability. 
There is a greater variation in the group-level parameters for same language, however (Figure~\ref{fig:posteriormugamma}, left panel), and 
$\Pr(\gamma^{[g]} >0 \mid y )$ ranges from $0.30$ to $1.00$ 
with a median of~$0.88$.


\end{document}